\documentclass[a4paper,11pt]{article}
\usepackage[dvipdfmx]{graphicx}
\usepackage{amssymb}
\usepackage{amsmath}
\usepackage{amsfonts}
\usepackage{amssymb}
\usepackage[nosort]{cite}
\usepackage{color}
\makeatletter
\renewcommand{\theequation}{\thesection.\arabic{equation}}
\newcommand{\qed}{\hbox{\rule[-2pt]{3pt}{6pt}}}
\newcommand{\mib}[1]{\mbox{\boldmath $#1$}}
\def\hs{\hspace{0.3 cm}}
\def\h{\hspace{0.22 cm}}
\def\hp{\hspace{0.29 cm}}
\def\hc{\hspace{0.218 cm}}
\def\he{\hspace{0.195 cm}}
\def\I{\boldsymbol{1}}
\@addtoreset{equation}{section}
\makeatother
\begin{document}
\title{\textbf{Oscillatory matrix model in Chern-Simons theory and
Jacobi-theta determinantal point process}}
\author{Yuta Takahashi \thanks{ytakahashi@phys.chuo-u.ac.jp}
and Makoto Katori\thanks{katori@phys.chuo-u.ac.jp} 
\and \textit{Department of Physics, Faculty of Science and Engineering,}
\and \textit{Chuo University, Kasuga, Bunkyo-ku, Tokyo 112-8551, Japan}}
\date{(12 Aug 2014)}
\maketitle
\begin{abstract}
The partition function of the Chern-Simons theory
on the three-sphere with the unitary group $U(N)$ 
provides a one-matrix model.
The corresponding $N$-particle system can be mapped to the
determinantal point process whose correlation kernel is expressed by
using the Stieltjes-Wigert orthogonal polynomials.
The matrix model and the point process are regarded as
$q$-extensions of the random matrix model in the Gaussian unitary ensemble
and its eigenvalue point process, respectively.
We prove the convergence of the $N$-particle system to an
infinite-dimensional determinantal point process in $N \to \infty$,
in which the correlation kernel is expressed by Jacobi's theta functions.
We show that the matrix model obtained by this limit realizes
the oscillatory matrix model in Chern-Simons theory discussed by
de Haro and Tierz.
\end{abstract}
\section{Introduction}
Chern-Simons theory on a three-manifold $M$
with a simply-laced gauge group $G$ is specified by the action
\begin{equation}
S(A) = \frac{k}{4 \pi} \int_M
\mathrm{Tr} \left( A \wedge dA + \frac{2}{3}
A \wedge A \wedge A \right),
\end{equation}
where $A$ is a $G$-connection on $M$ and $k$ is an integer.
The partition function of the Chern-Simons theory is then given by
\begin{equation}
Z_k\left(M,G\right) = \int \mathcal{D}A e^{iS(A)}
\end{equation}
with $i = \sqrt{-1}$.
Based on \cite{Wit89, Wit95}, Mari\~no showed in \cite{Marino} that
the partition function of Chern-Simons theory on Seifert spaces can
be calculated in a combinatorial way and expressed by
multiple integrals.\h In\h particular,\h when\h the\h gauge\h group\h $G$\h is\h chosen\h
as\h the\h unitary\h group\h $U(N)$, $N \in \{2,3,\cdots \}$,
the Chern-Simons partition function on the three-sphere $S^3$ is
expressed by \cite{Marino}
\begin{equation}
Z_k\left(S^3,U(N)\right)=\frac{e^{-g_s N(N^2-1)/12}}{N!} \int_{\mathbb{R}^N}
\prod_{j=1}^N \frac{d\phi_j}{2\pi} e^{-\phi_j^2/2g_s}
\prod_{1\leq j<k \leq N} \left(
2\sinh \frac{\phi_k - \phi_j}{2}\right)^2,
\label{csp}
\end{equation}
where the string coupling constant $g_s$ is given by
\begin{equation}
g_s = \frac{2\pi i }{ k+N}.
\label{gs}
\end{equation}
The structure of (\ref{csp}) is similar to
those of partition functions of one-matrix models
\cite{Mehta, Forrester}. Tierz \cite{Tierz2004} put
\begin{equation}
q=e^{-g_s}
\label{q}
\end{equation}
and regarded (\ref{csp}) as the partition function of matrix model
associated with the Stieltjes-Wigert polynomials
$p_n(\cdot;q)$, $n \in \mathbb{N}_0 = \{0,1,2,\ldots \}$,
which are $q$-extensions of Hermite polynomials (see Section 2.1).
He performed the integral (\ref{csp}) by using orthonormality of
$p_n(\cdot;q)$'s, which is generally
valid for $0<|q|<1$, and obtained the exact and explicit expression
for (\ref{csp})
\begin{equation}
Z_k\left(S^3,U(N)\right)=e^{i\pi N^2/4} (k+N)^{-N/2}
\prod_{j=1}^{N-1}\left( 2\sin \frac{\pi j}{k+N} \right)^{N-j}
\label{CSpart}
\end{equation}
by setting (\ref{gs}) in the result \cite{Tierz2004}.

The above fact leads us to consider the Chern-Simons partition
function (\ref{csp}) in the way that the constant $g_s$ is a free
parameter and is not restricted by (\ref{gs}).
In the present paper,
we consider the case that the parameter
$g_s$ is positive and regard
(\ref{csp}) as the partition function of
a statistical mechanics system of $N$ particles.
The variables $\phi_j \in \mathbb{R}$, $1 \leq j \leq N$
in the integral (\ref{csp}) are considered to be realizations of
$N$ random variables on $\mathbb{R}$ whose probability law
is given by the probability density function
\begin{equation}
\widetilde{\mathrm{P}}_N\left(\left\{ \phi_j \right\}_{j=1}^N\right)
=\widetilde{c}_N
\prod_{j=1}^N\frac{e^{-\phi_j^2/2g_s}}{\sqrt{2\pi g_s}}
\prod_{1\leq j<k \leq N} \left(
2\sinh\frac{\phi_k -\phi_j}{2} \right)^2.
\label{bm}
\end{equation}
Here $\widetilde{c}_N$ is the normalization constant
which will be explicitly given in Section 2.1.
From it, the partition function (\ref{csp}) is obtained by
\begin{equation}
Z_k\left(S^3,U(N)\right)=\left( \frac{g_s}{2\pi} \right)^{N/2}
\frac{e^{-g_sN(N^2-1)/12}}{ \widetilde{c}_N N!}.
\label{ZwhcN}
\end{equation}
Let $\mathbb{R}_{+} = \left\{ x \in \mathbb{R}: x >0 \right\}$.
By the mapping
\begin{equation}
f_N : \phi \in \mathbb{R} \mapsto x \in \mathbb{R}_{+},
\quad x = f_N(\phi) = e^{\phi+Ng_s},
\label{mapf}
\end{equation}
(\ref{bm}) is transformed to
the probability density function
\begin{equation}
\mathrm{P}_N\left(\left\{ x_j \right\}_{j=1}^N\right)
=c_N \prod_{j=1}^N \frac{e^{-(\ln x_j)^2/2g_s}}
{\sqrt{2\pi g_s}}
\prod_{1\leq j<k \leq N} (x_k - x_j)^2,
\label{SW}
\end{equation}
where $c_N$ is given by
\begin{equation}
c_N =e^{-g_s N^3 /2} \widetilde{c}_N.
\label{cNwhcN}
\end{equation}
In the Gaussian unitary ensemble (GUE) with variance $\sigma^2$,
the eigenvalues of $N \times N$ random Hermitian matrices obey
the probability density
\begin{equation}
\mathrm{P}^{\mathrm{GUE}}_N \left(\left\{ x_j \right\}_{j=1}^N\right)
=c^{\mathrm{GUE}}_N \prod_{j=1}^N \frac{e^{-x_j^2/2\sigma^2}}
{\sqrt{2\pi}\sigma} \prod_{1\leq j<k \leq N} (x_k - x_j)^2,
\end{equation}
where $c^{\mathrm{GUE}}_N = 1/\sigma^{N^2-N}\prod_{j=1}^N j!$
\cite{Mehta, Forrester}.
In the present ensemble (\ref{SW}), individual point $x_j$
follows the \textit{log-normal distribution} on $\mathbb{R}_+$
instead of the Gaussian distribution on $\mathbb{R}$,
while the repulsive interactions represented by
$\prod_{1\leq j<k \leq N} (x_k - x_j)^2$ are common.

Let $X_j$, $1 \leq j \leq N$ be random variables having the
probability density function (\ref{SW}) on $\mathbb{R}_+$.
Since $\mathrm{P}_N$ as well as $\mathrm{P}^{\mathrm{GUE}}_N$
are symmetric
functions of $x_j$, $1 \leq j \leq N$, we shall represent a
configuration as unlabeled. Let $\mathfrak{M}$ be the space
of nonnegative integer-valued Radon measure on $\mathbb{R}_+$.
Any element $\xi \in \mathfrak{M}$ is represented as
$\xi(\cdot)=\sum_{j \in \mathbb{I}} \delta_{x_j}(\cdot)$
with a countable index set $\mathbb{I}$, where
$\delta_{x}(\cdot)$ denotes a point mass (the delta measure)
on $x$. There a sequence of points in $\mathbb{R}_+$,
$\mib{x}=\left( x_j \right)_{j \in \mathbb{I}}$,
satisfies $\xi(K)=\sharp \left\{ x_j: x_j\in K \right\}<\infty$
for any compact subset $K \subset \mathbb{R}_+$.
Then we regard the present particle system as
$\mathfrak{M}$-valued and write it as $(\Xi, \mathrm{P}_N)$ with
\begin{equation}
\Xi(\cdot) = \sum_{j=1}^N \delta_{X_j}(\cdot).
\end{equation}

Let $\mathrm{C}_0 (\mathbb{R}_+)$ be the set of all continuous
real-valued function with compact support on $\mathbb{R}_+$.
For $f \in \mathrm{C}_0 (\mathbb{R}_+)$, the
\textit{moment generating function} of the system is given by
the following generalized
Laplace transform of the distribution (\ref{SW}),
\begin{eqnarray}
\mathcal{G}_N[f]&=&
\mathrm{E}_N \left[ \exp\left(
\int_{\mathbb{R}_+} f(x) \Xi(dx)
\right) \right]
\nonumber \\
&=& \int _{\mathbb{R}^N_+} d\mib{x}
\mathrm{P}_N\left(\left\{ x_j \right\}_{j=1}^N\right)
e^{\sum_{k=1}^N f(x_k)},
\label{mgf}
\end{eqnarray}
where $d\mib{x} = \prod_{j=1}^N dx_j$ and $\mathrm{E}_N$
denotes the expectation with respect to $\mathrm{P}_N$.
It is expanded with respect to the `test function'
$g(\cdot) = e^{f(\cdot)}-1$ as
\begin{eqnarray}
\mathcal{G}_N[f]
&=& \int_{\mathbb{R}^N_+} d\mib{x}
\mathrm{P}_N\left(\left\{ x_j \right\}_{j=1}^N\right)
\prod_{k=1}^N \big( 1+g(x_k) \big)
\nonumber \\
&=& 1+ \sum_{N'=1}^N \frac{1}{N'!}
\int_{\mathbb{R}^{N'}_+} d\mib{x}^{(N')}
\prod_{j=1}^{N'} g\left( x_{j}^{(N')} \right)
\rho_{N}^{(N')} \left( \mib{x}^{(N')} \right),
\end{eqnarray}
where $\mib{x}^{(N')}$ denotes
$(x_1^{(N')}, x_2^{(N')}, \ldots, x_{N'}^{(N')})$,
$1 \leq N' \leq N$.
Here $\rho_{N}^{(N')}$ gives the \textit{$N'$-point correlation function}
for $(\Xi, \mathrm{P}_N)$, which is a symmetric function,
$1 \leq N' \leq N$.
Given an integral kernel $\mib{K}(x,y)$,
$(x,y) \in \mathbb{R}^2_+$, a \textit{Fredholm determinant} with
$g \in \mathrm{C}_0 (\mathbb{R}_+)$ is defined as
\begin{eqnarray}
&&\mathop{\mathrm{Det}}_{(x,y)\in \mathbb{R}^2_+}
\Big[ \delta(x-y) +\mib{K}(x,y) g(y) \Big]
\nonumber \\
&&\quad\quad\quad
= 1 + \sum_{N'=1}^N \frac{1}{N'!}\int_{\mathbb{R}^{N'}_+} d\mib{x}^{(N')}
\prod_{j=1}^{N'} g\left( x_{j}^{(N')} \right)
\det_{1\leq k,\ell \leq N'}
\left[ \mib{K}\left(x_{k}^{(N')},x_{\ell}^{(N')} \right) \right].
\label{Fred}
\end{eqnarray}
If the system $(\Xi, \mathrm{P}_N)$ has an integral kernel $\mib{K}$ such that
any moment generating function (\ref{mgf}) is given by
a Fredholm determinant (\ref{Fred}),
$(\Xi, \mathrm{P}_N)$ is said to be a
\textit{determinantal point process with the correlation kernel}
$\mib{K}$ \cite{ST,Soshnikov}.
By definition, we have
\begin{eqnarray}
&&\rho_{N}^{(N')} \left( \mib{x}^{(N')} \right)
\nonumber \\
&&= \frac{N!}{(N-N')!} \int_{\mathbb{R}^{N-N'}_+} d\mib{x}^{(N-N')}
\mathrm{P}_N \left( \left\{
x_{1}^{(N')}, \ldots, x_{N'}^{(N')},
x_{1}^{(N-N')}, \ldots, x_{N-N'}^{(N-N')}
\right\} \right)
\label{def_of_cor} \\
&&= \det_{1\leq j,k \leq N'}
\left[ \mib{K}\left(x_{j}^{(N')},x_{k}^{(N')} \right) \right],
\label{correlationfunction}
\end{eqnarray}
$\mib{x}^{(N')} \in \mathbb{R}^{N'}_+$, $1 \leq N' \leq N$.
Note that the terminology `point process' does not mean any stochastic
process but does a spatial distribution of points as usually used in
probability theory (\textit{e.g.} Poisson processes).
Determinantal point process is also called fermion point process
\cite{Soshnikov, ST}.

As \cite{Tierz2004} implies and as a special case of result
in Section I\hspace{-.1em}I\hspace{-.1em}I.C
in our previous paper \cite{TK2012}, it is proved that the present system
$\left(\Xi,\mathrm{P}_N\right)$
is a determinantal point process with the correlation
kernel
\begin{eqnarray}
K_N (x,y) &=& \sum_{n =0}^{N-1} p_{n}(x;q) p_{n}(y;q)\sqrt{w(x;q)w(y;q)}
\nonumber \\
&=& \frac{\sqrt{1-q^N}}{q^{2N}}
\frac{p_N(x;q)p_{N-1}(y;q)-p_N(y;q)p_{N-1}(x;q)}{x-y}
\sqrt{w(x;q)w(y;q)},
\nonumber \\
\label{cd}
&&\quad\quad\quad\quad\quad\quad\quad\quad\quad\quad\quad\quad
\quad\quad\quad\quad\quad\quad\quad\quad\quad
\quad \mbox{$(x,y) \in \mathbb{R}^2_+$, $x \neq y$},
\\
K_N (x,x) &=& \sum_{n =0}^{N-1} p_{n}(x;q)^2 w(x;q)
\nonumber \\
&=& \frac{\sqrt{1-q^N}}{q^{2N}} \Big\{
p'_N(x;q)p_{N-1}(x;q)-p_N(x;q)p'_{N-1}(x;q) \Big\} w(x;q),
\hspace{0.3cm} \mbox{$x \in \mathbb{R}_+$},
\label{cd_onep}
\end{eqnarray}
where
$p_n(\cdot;q)$, $n \in \mathbb{N}_0$,
$0<q<1$ are the
Stieltjes-Wigert polynomials,
$p'_n(\cdot;q)$, $n \in \mathbb{N}_0$ are their derivatives,
and $w(\cdot;q)$ is their weight function for orthogonality,
which will be explicitly given in Section 2.1.
The second equality in (\ref{cd}) is given by
the Christoffel-Darboux formula \cite{Szego}.

This fact implies that the original system
\begin{equation}
\widetilde{\Xi}(\cdot) = \sum_{j=1}^N \delta_{\phi_j}(\cdot)
\end{equation}
with the probability density (\ref{bm})
associated with the Chern-Simons partition function
is also a determinantal point process
on $\mathbb{R}$. By (\ref{mapf}) the correlation kernel
of $(\widetilde{\Xi}, \widetilde{\mathrm{P}}_N)$ is given by
\begin{equation}
\mathcal{K}_N \left(\phi,\varphi\right) = e^{(\phi+\varphi)/2 + Ng_s}
K_N\left(e^{\phi + Ng_s},e^{\varphi + Ng_s}\right),
\quad \mbox{$(\phi, \varphi) \in \mathbb{R}^2$}.
\label{CSNkernel}
\end{equation}
These finite point processes $\left(\Xi,\mathrm{P}_N\right)$
and $(\widetilde{\Xi}, \widetilde{\mathrm{P}}_N)$
are fully studied
in \cite{Tierz2004, Tierz2005, DT2007, TK2012}.

The purpose of the present paper is to consider an $N \to \infty$
limit of the systems $(\Xi, \mathrm{P}_N)$ 
and $(\widetilde{\Xi}, \widetilde{\mathrm{P}}_N)$.
For $n \in \mathbb{N}_0$, let
\begin{eqnarray}
\chi(n)=\begin{cases}
0 &\mbox{if $n$ is even,}\\
1 &\mbox{if $n$ is odd,}
\end{cases}
\label{chi}
\end{eqnarray}
and for $x \in \mathbb{R}$, let $\lceil x \rceil$ be the least
integer not less than $x$.
Then
for $0<q<1$ and $2/3 < \tau < 2$, we will prove
\begin{eqnarray}
q^{-\lceil \tau N \rceil -\chi \left( \lceil \tau N \rceil \right)}
K_N\left(q^{-\lceil \tau N \rceil -\chi \left( \lceil \tau N \rceil \right)}u,
q^{-\lceil \tau N \rceil -\chi \left( \lceil \tau N \rceil \right)}v \right)
\to K^{\Theta}(u,v),
\quad \mbox{as $N \overset{\tau}{\to} \infty$},
\label{limitksw}
\end{eqnarray}
$(u,v) \in \mathbb{R}_+^2$
(Proposition 3),
where
\begin{eqnarray}
&&K^{\Theta}(u,v)= 
\frac{\displaystyle{  \Theta \big( -q^{1/2}u \big|q\big) \Theta (-q^{-1/2}v |q)
- \Theta \big(-q^{1/2}v\big|q\big) \Theta \big(-q^{-1/2}u\big|q\big) } }
{\displaystyle{u-v}}
\frac{\sqrt{w(u;q)w(v;q)}}{\prod_{k=1}^{\infty}(1-q^{k})^3},
\nonumber \\
&&\quad\quad\quad\quad\quad\quad\quad
\quad\quad\quad\quad\quad\quad\quad\quad\quad\quad
\quad\quad\quad\quad\quad\quad\quad\quad\quad\quad
\quad\mbox{$(u,v) \in \mathbb{R}_+^2$, $u \neq v$},
\label{thetakernel}
\end{eqnarray}
\begin{eqnarray}
&&\hspace{-0.5cm}K^{\Theta}(u,u)=
\left\{
\frac{1}{\sqrt{q}}\Theta(-q^{1/2}u|q) \Theta'(-q^{-1/2}u |q)
- \sqrt{q}\Theta' (-q^{1/2}u|q) \Theta ( -q^{-1/2}u |q)
\right\}
\frac{w(u;q)}{\prod_{k=1}^{\infty}(1-q^{k})^3},
\nonumber \\
&&\quad\quad\quad\quad\quad\quad\quad\quad\quad\quad\quad
\quad\quad\quad\quad\quad\quad\quad\quad\quad\quad
\quad\quad\quad\quad\quad\quad\quad\quad\quad\quad\quad
\quad\mbox{$u \in \mathbb{R}_+$},
\label{thetakernel_onep}
\end{eqnarray}
with the weight function $w(\cdot;q)$,
and the conditional limit depending on $\tau$,
denoted by $N \overset{\tau}{\to} \infty$, is defined at the beginning of
Section 3.1.
Here $\Theta (\cdot|q)$ is a version of
Jacobi's theta function defined by
(see, for instance, \cite{WW}),
\begin {eqnarray}
\Theta(z|q)&=&\sum_{k=-\infty}^{\infty} q^{k^2} z^k
\label{thetafnc}
\\
&=& \prod_{k=1}^{\infty}(1-q^{2k})(1+zq^{2k-1})(1+z^{-1}q^{2k-1}),
\label{tz}
\end{eqnarray}
for $0<|q|<1$ and $0<|z|<\infty$,
and $\Theta' (\cdot|q)$ is its derivative
\begin{equation}
\Theta'(z|q) = \frac{d\Theta(z|q)}{dz}
=\frac{1}{z} \sum_{k=-\infty}^{\infty} k q^{k^2} z^k.
\end{equation}
Thus we call the correlation kernel $K^{\Theta}$ given by
(\ref{thetakernel}) and (\ref{thetakernel_onep})
the \textit{Jacobi-theta kernel}.
The convergence (\ref{limitksw}) of correlation kernel in $N \to \infty$
implies that of moment generating function $\mathcal{G}_N$
to the Fredholm determinant associated with the Jacobi-theta kernel
(\ref{thetakernel}), (\ref{thetakernel_onep}),
\begin{equation}
\mathop{\mathrm{Det}}_{(u,v)\in \mathbb{R}^2_+}
\Big[ \delta(u-v) +K^{\Theta}(u,v) g(v) \Big],
\label{Fredt}
\end{equation}
$g \in \mathrm{C}_0(\mathbb{R}_+)$.
Then all correlation functions $\rho_{N'}$, $N' \in \mathbb{N}$ are
determined and expressed by determinants.
In this sense, as the $N \to \infty$ limit of $(\Xi,\mathrm{P}_N)$
and $(\widetilde{\Xi},\widetilde{\mathrm{P}}_N)$,
determinantal point processes with infinite numbers of
particles are obtained (Theorem 4 and Corollary 5).

In \cite{Tierz2005} de Haro and Tierz discussed an
\textit{oscillatory matrix model},
which seems to appear in sufficiently large but finite $N$ in the
Chern-Simons theory with the $U(N)$ gauge. With the restriction (\ref{gs}),
the $N \to \infty$ limit is identified with the 't Hooft limit;
$N \to \infty$  with $Ng_s = \mathrm{constant}$.
Since it corresponds to $q \to 1$ by (\ref{q}), the oscillatory behavior
will vanish and only classical matrix model 
is obtained in the limit.
The oscillatory matrix model of de Haro and Tierz is realized as
a \textit{crossover phenomenon} in \cite{Tierz2005}.
In the present paper, we fix $g_s$ so that $0<q=e^{-g_s}<1$ and take limit
$N \to \infty$. The infinite-dimensional determinantal point process obtained
by this limit of $(\widetilde{\Xi},\widetilde{\mathrm{P}}_N)$ will be a
\textit{stationary} realization of the oscillatory matrix model observed
by de Haro and Tierz \cite{Tierz2005}. The oscillatory behavior will be
demonstrated in Section 4.1 with figures.
In Section 4.2 we will confirm that
if we take the further limit $g_s \to 0$ (\textit{i.e.} $q \to 1$), the system
becomes classical with the sine-kernel as it should.
In other words, in the context of random matrix theory \cite{Mehta, Forrester}
the present paper reports a $q$-extension of the bulk scaling limit of the
Hermite kernel in GUE. The $q$-extension of the edge scaling limit
described by the Airy kernel will be reported in a forthcoming paper \cite{preparation}.

The paper is organized as follows.
In Section 2, we define the Stieltjes-Wigert polynomials $p_n$
and give their asymptotic expansions as the degree of polynomials
$n \to \infty$.
In Section 3, we present asymptotic form of the Stieltjes-Wigert
kernel described by (\ref{cd}) and (\ref{cd_onep}) and explain its connection
with the oscillatory matrix model.
Section 4 is devoted to showing the oscillatory behaviors
of the infinite-particle systems.
Proofs of Lemma 1 and Proposition 3 are given in Section 5.
In Appendix A, we rewrite the correlation kernel of the oscillatory
matrix model in standard notations of theta functions \cite{WW}
and in terms of Gosper's $q$-trigonometric functions \cite{Gosper}.
Appendix B complements
the proof of Lemma 1.
\section{Preliminaries}
\subsection{Some $q$-special functions}
For $0<|q|<1$ and $z \in \mathbb{C}$,
we introduce the $q$-Pochhammer symbol
\begin{equation}
(z;q)_0 = 1, \quad
(z;q)_n = \prod_{k=0}^{n-1} (1-zq^k),
\quad \mbox{$n \in \mathbb{N}$},
\end{equation}
and
\begin{equation}
(z;q)_{\infty} = \lim_{n \to \infty} (z;q)_n.
\end{equation}
The following identity follows from the $q$-binomial theorem \cite{AAR},
\begin{eqnarray}
(z;q)_n
= \sum_{k=0}^n \frac{(q;q)_n q^{k(k-1)/2}(-z)^k}
{(q;q)_k (q;q)_{n-k}},
\quad \mbox{$n \geq 0$},
\end{eqnarray}
and then
\begin{equation}
(z;q)_{\infty}
= \sum_{k=0}^{\infty} \frac{q^{k(k-1)/2}(-z)^k}
{(q;q)_k}.
\label{qPs}
\end{equation}

For $0<q <1$ and
$n \in \mathbb{N}_0$, the orthonormal
Stieltjes-Wigert polynomials are defined by \cite{Szego}
\begin{equation}
p_n(x;q) = (-1)^n q^{n/2+1/4} \sqrt{(q;q)_n}
\sum_{k=0}^n \frac{q^{k^2}(-q^{1/2}x)^k}{(q;q)_k (q;q)_{n-k}},
\quad \mbox{$x>0$}.
\label{onSWpoly}
\end{equation}
They satisfy the orthonormality relations
\begin{equation}
\int_0^{\infty} p_n(x;q) p_m(x;q) w(x;q)dx
= \delta_{nm}, \quad \mbox{$n,m \in \mathbb{N}_0$},
\label{onorm}
\end{equation}
with respect to the weight function
\begin{equation}
w(x;q)=\frac{1}{\sqrt{2\pi |\ln q|}}
\exp \left[ -\frac{(\ln x)^2}{2|\ln q|} \right],
\quad \mbox{$x >0$}.
\label{SWwei}
\end{equation}
This gives a density for a log-normal distribution
and solves the functional equation
\begin{equation}
w(q^{s}x;q)=q^{s^2/2} x^{s} w(x;q),
\label{weq}
\quad s\in \mathbb{R}.
\end{equation}
By using the Stieltjes-Wigert polynomials and their orthonormality
(\ref{onorm}),
normalization constant $c_N$ in (\ref{SW}) is determined as
\begin{equation}
c_N = \frac{e^{-g_s N(4N^2-1)/6}}
{N!\prod_{k=1}^{N-1}(e^{-g_s};e^{-g_s})_k},
\end{equation}
where (\ref{q}) was assumed.
Then $\widetilde{c}_N$ is given by (\ref{cNwhcN}),
and through (\ref{ZwhcN}) we obtain
\begin{equation}
Z_k\left( S^3,U(N) \right) = \left( \frac{g_s}{2\pi} \right)^{N/2}
e^{g_s N(N^2-1)/12}
\prod_{j=1}^{N-1}(e^{-g_s};e^{-g_s})_j.
\label{matrixZ}
\end{equation}
We have the identity
\begin{equation}
\prod_{j=1}^{N-1}(e^{-g_s};e^{-g_s})_j
=e^{-g_s N(N^2-1)/12} e^{i \pi N(N-1)/4}
\prod_{j=1}^{N-1}\left( 2\sin \frac{jg_s}{2i} \right)^{N-j},
\end{equation}
and obtain the expression (\ref{CSpart})
by substituting (\ref{gs}) into (\ref{matrixZ}) \cite{Tierz2004}.

The theta function (\ref{tz}) can be written as
\begin {eqnarray}
\Theta(z|q)
=(q^2;q^2)_{\infty} (-zq;q^2)_{\infty} (-z^{-1}q;q^2)_{\infty},
\label{J3}
\end{eqnarray}
which is called Jacobi's triple product identity.
One can prove the functional equation
\begin{equation}
\Theta(q^2z|q) = q^{-1}z^{-1}\Theta(z|q),
\label{fnceqT}
\end{equation}
directly from the definition (\ref{thetafnc}).

A $q$-exponential function is defined as
\begin{equation}
e_q(z) = (z;q)_{\infty}^{-1},\quad |z|<1.
\label{eq}
\end{equation}
\subsection{Asymptotic expansions for the Stieltjes-Wigert polynomials}
For $\tau \in (0,2)$, $n \in \mathbb{N}$, let
\begin{equation}
m = \lfloor (2-\tau)n \rfloor,
\label{m}
\end{equation}
and
\begin{equation}
\lambda = (2-\tau)n -m,
\label{lambda}
\end{equation}
where $\lfloor x \rfloor$ denotes the integer part of $x \in \mathbb{R}$.
We also introduce
the indicator function $\I_A(\omega)$ of a set $A$ such that
$\I_A(\omega)=1$ if $\omega \in A$ and $\I_A(\omega)=0$ otherwise.
\vskip 0.5cm
\noindent \textbf{Lemma 1} \quad
Let $0<q<1$ and $0<\tau<2$. Then
the orthonormal Stieltjes-Wigert polynomials (\ref{onSWpoly}) have
the following asymptotic expansions as
the degree of polynomials $n \to \infty$,
\begin{eqnarray}
&&p_n \left(q^{-n\tau}u;q \right)\nonumber \\
&&\quad=\frac{(-1)^n q^{n/2+1/4 +n^2 (1-\tau)
-\lfloor m/2\rfloor (\lfloor m/2\rfloor + \chi(m) + \lambda)}
\sqrt{(q;q)_n}}
{(-q^{1/2}u)^{\lfloor m/2 \rfloor -n} (q;q)_{\infty}^2}
\bigg\{ \Theta\left(-q^{\lambda+\chi(m)+1/2} u\Big|q\right)
\nonumber \\
&&\qquad+\frac{q^{\lfloor m/2 \rfloor}}{1-q}
\left( \frac{q^{1/2-(1-\tau)n}}{u}\I_{(0,4/3)}(\tau)
- q\I_{(2/3,2)}(\tau) \right)
\Theta\left(-q^{\lambda+\chi(m)-1/2} u\Big|q\right)
\nonumber\\
&&\quad\quad\quad\quad\quad\quad\quad\quad\quad\quad\quad
\quad\quad\quad\quad\quad\quad\quad\quad\quad\quad\quad\quad
+{\cal O}\left(q^{\tau n +2(1-\tau)n \I_{[1,2)}(\tau)}\right) \bigg\}.
\label{onSWayex}
\end{eqnarray}
\vskip 0.5 cm
\noindent
The proof is given in Section 5.1 with Appendix B.
Since the Stieltjes-Wigert polynomial is a $q$-extension
of the Hermite polynomials \cite{Koekoe}, this result can be
regarded as a $q$-analogue of the celebrated
Plancherel-Rotach asymptotic formula \cite{PR}.
We note that the leading term given by the first term
in the parenthesis in the RHS was
given by Ismail and Zhang
as equations (2.19) and (2.23) in \cite{IsmailZhang}
(and (16) and (19) in \cite{IsmailZhang2}).
This lemma improves their estimate.
In order to obtain the Jacobi-theta kernel given by
(\ref{thetakernel}) and (\ref{thetakernel_onep})
as a limit of the Stieltjes-Wigert kernel
expressed by (\ref{cd}) and (\ref{cd_onep}),
the correction term given by
the second term in the parenthesis
is necessary (Proposition 3). 
Owing to the factor $\lfloor m/2 \rfloor$ with (\ref{m}) in the
formula, asymptotic behavior in $n \to \infty$
will depend on whether $\tau$ is rational or irrational
as discussed in \cite{IsmailZhang}.

Lemma 1 gives the following asymptotic expansions for
$p_n (q^{-n\tau}u;q)$ multiplied by
the weight function (\ref{SWwei}); as $n \to \infty$,
\noindent
\begin{eqnarray}
&&p_n \left(q^{-n\tau}u;q\right)
\sqrt{w\left(q^{-n\tau}u;q\right)}\nonumber \\
&&\quad=\frac{ \sqrt{(q;q)_n} }{(q;q)_{\infty}^2}
(-1)^{\lfloor m/2 \rfloor}
q^{n+1/4-\lfloor m/2 \rfloor/2}
\sqrt{w\left(q^{\lambda + \chi(m)}u;q\right)}
\bigg\{ \Theta\left(-q^{\lambda+\chi(m)+1/2} u\Big|q\right)
\nonumber \\
&&\qquad
+\frac{q^{\lfloor m/2 \rfloor}}{1-q}
\left( \frac{q^{1/2-(1-\tau)n}}{u}\I_{(0,4/3)}(\tau)
- q\I_{(2/3,2)}(\tau) \right)
\Theta\left(-q^{\lambda+\chi(m)-1/2} u\Big|q\right)
\nonumber\\
&&\quad\quad\quad\quad\quad\quad\quad
\quad\quad\quad\quad\quad\quad\quad\quad
\quad\quad\quad\quad\quad\quad\quad\quad
+{\cal O}\left(q^{\tau n +2(1-\tau)n \I_{[1,2)}(\tau)}\right) \bigg\}.
\end{eqnarray}
\noindent
Then, we can find the following.
\vskip 0.5cm
\noindent \textbf{Lemma 2} \quad
For $0<q<1$, $0<\tau<2$,
\begin{eqnarray}
&&\hspace{-0.5cm}p_n \left(q^{-\lceil \tau n \rceil 
-\chi \left(\lceil \tau n \rceil \right)}u;q \right)
\sqrt{w\left(q^{-\lceil \tau n \rceil 
-\chi \left(\lceil \tau n \rceil \right)}u;q \right)}
\nonumber \\
&&\hspace{-0.5cm}\quad=
\frac{ \sqrt{(q;q)_n} }{(q;q)_{\infty}^2} (-1)^{n}
q^{n/2+1/4+\lceil \tau n \rceil /4
+ \chi \left(\lceil \tau n \rceil \right)/4}
\sqrt{w(u;q)}
\bigg\{ \Theta\left(-q^{1/2} u\Big|q\right) \nonumber \\
&&\hspace{-0.5cm}\qquad
+\frac{q^{n - \lceil \tau n \rceil /2
- \chi \left(\lceil \tau n \rceil \right)/2}}
{1-q}
\left( \frac{q^{1/2-n+ \lceil \tau n\rceil
+\chi \left(\lceil \tau n \rceil \right)}}
{u} \I_{(0,4/3)}(\tau)- q\I_{(2/3,2)}(\tau) \right)
\Theta\left(-q^{-1/2} u\Big|q\right)
\nonumber\\
&&\hspace{-0.5cm}\quad\quad\quad\quad\quad\quad\quad\quad\quad
\quad\quad\quad\quad\quad\quad\quad\quad\quad\quad
+{\cal O}\left(q^{\tau n +2(1-\tau)n \I_{[1,2)}(\tau)}\right) \bigg\},
\quad
\mbox{as $n \to \infty$.}
\label{onSWayexww}
\end{eqnarray}
\section{Main theorems}
\subsection{Jacobi-theta determinantal point process}
Since we obtained the asymptotic forms of the Stieltjes-Wigert
polynomials as Lemmas 1 and 2,
we will be able to determine the asymptotics of
the Stieltjes-Wigert kernel given by (\ref{cd}) and (\ref{cd_onep})
in $N \to \infty$.
For a technical reason, here we assume $\tau \in (2/3,2)$.
We consider a monotonically increasing series
of integers $(N_j)_{j \in \mathbb{N}}$ such that the equalities
\begin{eqnarray}
\lceil \tau(N_j-1) \rceil = \lceil \tau N_j \rceil + \lceil -\tau \rceil,
\quad \mbox{$j \in \mathbb{N}$},
\label{assumption}
\end{eqnarray}
holds. Note that $\lceil -\tau \rceil = -\I_{[1,2)}(\tau)$
for $\tau \in (2/3,2)$.
Then $(N_j^{(0)})_{j \in \mathbb{N}}$ and $(N_j^{(1)})_{j \in \mathbb{N}}$
are defined as subsequences of $(N_j)_{j \in \mathbb{N}}$ such that
$\lceil \tau N_j^{(0)} \rceil$ are even and
$\lceil \tau N_j^{(1)} \rceil$ are odd, respectively, $j \in \mathbb{N}$.
For a given $\tau \in (2/3,2)$, we take the limit $N \to \infty$
following the subsequences $(N_j^{(0)})_{j \in \mathbb{N}}$ and
$(N_j^{(1)})_{j \in \mathbb{N}}$.
We write this conditional limit as $N \overset{\tau}{\to} \infty$.

We have the following result.
\vskip 0.5cm
\noindent \textbf{Proposition 3} \quad
Let $(u,v)\in \mathbb{R}_+^2$ and $2/3 < \tau <2$.
Then (\ref{limitksw}) holds.
\vskip 0.5 cm
\noindent
We expect that the statement will be extended for $\tau \in (0,2/3]$,
but we need further improvement of Lemmas 1 and 2 to prove it.

Proposition 3 means the convergence of integral operators
\begin{eqnarray}
\widehat{K}_N f(\cdot) &\equiv& \int_{\mathbb{R}_+} dv
q^{-\lceil \tau N \rceil
-\chi \left( \lceil \tau N \rceil \right)}
K_N\left(q^{-\lceil \tau N \rceil
-\chi \left( \lceil \tau N \rceil \right)}\cdot,
q^{-\lceil \tau N \rceil
-\chi \left( \lceil \tau N \rceil \right)}v \right)f(v)
\nonumber \\
&\longrightarrow&
\widehat{K}^{\Theta} f(\cdot) = \int_{\mathbb{R}_+} dv
K^{\Theta}(\cdot,v) f(v),
\quad\mbox{in $N \overset{\tau}{\to} \infty$},
\end{eqnarray}
$f \in \mathrm{C}_0(\mathbb{R}_+)$.
The convergence of integral operators
$\widehat{K}_N \to \widehat{K}^{\Theta}$ implies
that of Fredholm determinants to (\ref{Fredt})
for $g \in \mathrm{C}_0(\mathbb{R}_+)$.
Since the Fredholm determinants are identified with
the moment generating functions in determinantal point process,
we can conclude the following.
\vskip 0.5cm
\noindent \textbf{Theorem 4} \quad
Let $2/3 < \tau <2$.
The determinantal point process $(\Xi,\mathrm{P}_N)$
converges to the determinantal point process in
$N \overset{\tau}{\to} \infty$, whose
correlation kernel is given by the Jacobi-theta kernel
$K^{\Theta}$ defined by (\ref{thetakernel}) and
(\ref{thetakernel_onep}).
In other words, for any $N' \in \mathbb{N}$,
\begin{eqnarray}
\rho_{N}^{(N')} 
\left(q^{-\lceil \tau N \rceil
-\chi \left( \lceil \tau N \rceil \right)}
\mib{x}^{(N')}\right)
\to \rho^{(N')} \left(\mib{x}^{(N')}\right)
=\det_{1\leq j,k\leq N'}
\bigg[ K^{\Theta} \left(x_j^{(N')},x_k^{(N')} \right) \bigg],
\quad \mbox{as $N \overset{\tau}{\to} \infty$},
\end{eqnarray}
where $q^{-\lceil \tau N \rceil
-\chi \left( \lceil \tau N \rceil \right)}
\mib{x}^{(N')}
\equiv \big( q^{-\lceil \tau N \rceil
-\chi \left( \lceil \tau N \rceil \right)}
x_j^{(N')} \big)_{j=1}^{N'}$.
\subsection{Mapping to the matrix model}
We obtained an infinite-particle system on $\mathbb{R}_+$ in the
previous subsection. Here, we
explain that the particle system is
then mapped to an infinite-particle system on
$\mathbb{R}$, which will be regarded as a stationary
realization of the oscillatory matrix model
considered in \cite{Tierz2005}.

First, we remind that the $N$-particle systems
$(\widetilde{\Xi},\widetilde{\mathrm{P}}_N)$
and $(\Xi,\mathrm{P}_N)$ were related
by the mapping (\ref{mapf}).
On the other hand, when we take the
$N \to \infty$ limit in Theorem 4,
we performed the scaling of variables as
\begin{equation}
h_{N,\tau} : x \in \mathbb{R}_{+} \mapsto u \in \mathbb{R}_{+},
\quad u = h_{N,\tau}(x) =
q^{\lceil \tau N \rceil +\chi\left( \lceil \tau N \rceil \right)}x,
\label{maph}
\end{equation}
where $q=e^{-g_s}$.
Then, the combination of the mappings (\ref{mapf}) and
(\ref{maph}) gives
\begin{equation}
h_{N,\tau} \circ f_N : \phi \in \mathbb{R} \mapsto u \in \mathbb{R}_{+},
\quad u = h_{N,\tau} \circ f_N(\phi) = e^{\phi + g_s\left(
N - \lceil \tau N \rceil -\chi\left( \lceil \tau N \rceil \right) \right)}.
\label{comp}
\end{equation}
\vskip 0.1cm
\noindent
This suggests that, only in the case of $\tau=1$, the $N$-dependent factor
$N - \lceil \tau N \rceil$ vanishes. 
(The value of the factor $\chi\left( \lceil \tau N \rceil \right)$ is fixed
to be $0$ in the series $(N_j^{(0)})_{j \in \mathbb{N}}$ and $1$
in $(N_j^{(1)})_{j \in \mathbb{N}}$ by the definition of $\chi$, (\ref{chi}),
and of $N \overset{\tau}{\to} \infty$.)
Hence, in the case $\tau = 1$, if we take an infinite-particle
limit of the scaled version of $(\Xi,\mathrm{P}_N)$ by (\ref{maph}),
we can obtain an infinite-dimensional model in Chern-Simons theory
by simply putting $u=e^{\phi - g_s \chi(N)}$.

Let
\begin{eqnarray}
\mathcal{K}_{\infty}\left(\phi,\varphi\right)
&=&e^{(\phi+\varphi)/2} K^{\Theta}\big(e^{\phi}, e^{\varphi}\big)
\nonumber \\
&=&\frac{e^{-(\phi^2+\varphi^2)/4g_s} / \sqrt{2\pi g_s}}
{(e^{-g_s};e^{-g_s})_{\infty}^{3} }
\nonumber \\
&&\times
\frac{\Theta\left(-e^{\phi-g_s/2}\big|e^{-g_s}\right)
\Theta\left(-e^{\varphi+g_s/2}\big|e^{-g_s}\right)
-\Theta\left(-e^{\varphi-g_s/2}\big|e^{-g_s}\right)
\Theta\left(-e^{\phi+g_s/2}\big|e^{-g_s}\right) }
{\displaystyle{2\sinh \frac{\phi-\varphi}{2}}},
\nonumber \\
&&\qquad\qquad\qquad\qquad\qquad\qquad\qquad\qquad
\qquad\qquad\qquad\qquad
\mbox{$(\phi,\varphi) \in \mathbb{R}^2$, $\phi \neq \varphi$,}
\label{bmkernel}
\end{eqnarray}
and
\begin{eqnarray}
\mathcal{K}_{\infty}(\phi,\phi)
&=&e^{\phi} K^{\Theta}\big(e^{\phi}, e^{\phi}\big)
\nonumber \\
&=&\frac{e^{-(\phi-g_s)^2/2g_s} / \sqrt{2\pi g_s}}
{(e^{-g_s};e^{-g_s})_{\infty}^{3} }
\nonumber \\
&&\times
\left\{ e^{g_s}
\Theta(-e^{\phi-g_s/2}|e^{-g_s})\Theta'(-e^{\phi+g_s/2}|e^{-g_s})
-\Theta'(-e^{\phi-g_s/2}|e^{-g_s})\Theta(-e^{\phi+g_s/2}|e^{-g_s})
\right\},
\nonumber \\
&&\quad\quad\quad\quad\quad\quad\quad\quad
\quad\quad\quad\quad\quad\quad\quad\quad
\quad\quad\quad\quad\quad\quad\quad
\quad\quad\quad\quad\quad\quad
\quad \mbox{$\phi \in \mathbb{R}$}.
\label{bmkernel_onep}
\end{eqnarray}
Then, for $(\widetilde{\Xi},\widetilde{\mathrm{P}}_N)$
with correlation kernel (\ref{CSNkernel}),
we have the following corollary from Proposition 3 and Theorem 4.
\vskip 0.5cm
\noindent \textbf{Corollary 5} \quad
Set $\tau =1$. Then
\begin{eqnarray}
\mathcal{K}_{N}\left(\phi+g_s\chi(N),\varphi+g_s\chi(N)\right)
\to \mathcal{K}_{\infty}(\phi,\varphi),
\quad \mbox{as $N \overset{\tau=1}{\longrightarrow} \infty$}.
\label{wsl}
\end{eqnarray}
Then, the system $(\widetilde{\Xi},\widetilde{\mathrm{P}}_N)$
converges to an infinite-dimensional determinantal point process
on $\mathbb{R}$ with the correlation kernel
$\mathcal{K}_{\infty}$ in $N \overset{\tau=1}{\longrightarrow} \infty$.
The correlation functions are given by
\begin{equation}
\widetilde{\rho}^{(N')}\left( \boldsymbol{\phi}^{(N')} \right)
= \det_{1\leq j,k \leq N'} \left[ \mathcal{K}_{\infty}
\left( \phi_j^{(N')} , \phi_k^{(N')}\right) \right],
\quad \mbox{$N' \in \mathbb{N}$},
\end{equation}
for the limit system, where $\boldsymbol{\phi}^{(N')}$ denotes
$(\phi_1^{(N')}, \phi_2^{(N')},
\ldots, \phi_{N'}^{(N')}) \in \mathbb{R}^{N'}$.
\vskip 0.5cm

When $\tau =1$, the condition (\ref{assumption}) is always satisfied;
$\lceil N_j -1 \rceil = \lceil N_j \rceil -1$.
Then the limit $N \overset{\tau=1}{\longrightarrow} \infty$ is just a
conditional limit such that we consider the limit in the even numbers
$N_j^{(0)} = 2j$ and in the odd numbers $N_j^{(1)} = 2j+1$,
$j \in \mathbb{N}$, separately.

We note that in the infinite-particle limit
$N \overset{\tau=1}{\longrightarrow} \infty$ in Corollary 5,
we do not need any scaling,
while we do in the bulk scaling limit of the
GUE-determinantal point process \cite{Mehta, Forrester}.
The situation seems to be quite similar to the $N \to \infty$
limit of the Ginibre determinantal point process on
$\mathbb{C} \simeq \mathbb{R}^2$ \cite{Ginibre}.

Other expressions of $\mathcal{K}_{\infty}$
by using Jacobi's theta functions $\vartheta_j$, $1\leq j \leq 4$
\cite{WW} and Gosper's $q$-trigonometric
functions \cite{Gosper} are given in Appendix A.

In the present paper, we consider the determinantal point processes
on $\mathbb{R}$ and $\mathbb{R}_+$, which are mapped by (\ref{mapf})
to each other. It will be an interesting future problem to study the
matrix model and the associated point process on the unit circle, 
which will be transformed from the present systems as discussed
in \cite{Oku05,OSY11,RT12,SzaboTierz2012}.

\section{Oscillatory matrix model in Chern-Simons theory}
\subsection{Oscillatory behavior}
From the pseudo-periodicity of Jacobi's theta function (\ref{fnceqT})
we have
\begin{equation}
\Theta ( -e^{(\phi +2g_s) \pm g_s/2 } | e^{-g_s} )
= - e^{g_s} e^{\phi \pm g_s/2}
\Theta ( -e^{\phi \pm g_s/2} | e^{-g_s} )
\label{pseudop}
\end{equation}
for theta functions used to express $\mathcal{K}_{\infty}$ in
(\ref{bmkernel}) and (\ref{bmkernel_onep}).
Then
we can readily prove the periodicity of the correlation kernel
$\mathcal{K}_{\infty}$,
\begin{equation}
\mathcal{K}_{\infty} (\phi+2ng_s,\varphi+2ng_s)
= \mathcal{K}_{\infty} (\phi,\varphi ),
\quad \mbox{$(\phi, \varphi)\in \mathbb{R}^2$, $n \in \mathbb{Z}$}.
\label{perio}
\end{equation}
Note that the Jacobi-theta kernel $K^{\Theta}$ given by (\ref{thetakernel})
and (\ref{thetakernel_onep}) does not have such periodicity and
only has the quasi-periodicity,
\begin{equation}
q^{2n}K^{\Theta}(q^{2n}u,q^{2n}v)=K^{\Theta}(u,v),
\quad \mbox{$(u,v)\in \mathbb{R}^2_{+}$, $0<q<1$, $n \in \mathbb{Z}$}.
\end{equation}

Then
all correlation functions
of the infinite-particle system on $\mathbb{R}$ obtained in Corollary 5
have oscillatory behavior caused by
(\ref{perio}). In order to demonstrate it, we consider
the case $N' =1$ here. In this case, (\ref{correlationfunction})
gives the density of the number of particles
\begin{equation}
\widetilde{\rho}(\phi) = \widetilde{\rho}^{(1)}_{\infty}(\phi)
= \mathcal{K}_{\infty}(\phi,\phi), \quad \mbox{$\phi \in \mathbb{R}$},
\label{rho}
\end{equation}
which is given by (\ref{bmkernel_onep}).
We can prove
\begin{equation}
\widetilde{\rho}(\phi+2ng_s)=
\widetilde{\rho}(\phi),
\quad \mbox{$\phi\in \mathbb{R}$, $n \in \mathbb{Z}$}.
\label{rhoperi}
\end{equation}
Figures 1-3 show the density profiles
(\ref{rho}) for $g_s=1$, $5$, and $25$, respectively.
We can see that as the period $2g_s$ increases,
both of the mean value of $\widetilde{\rho}(\phi)$
and the amplitude of oscillation decrease.
The mean value of $\widetilde{\rho}(\phi)$ can be estimated
by a `mean-field' analysis.
Mari\~no computed a density profile of the system
($\Xi, \mathrm{P}_N)$ by this approximation \cite{Marino_book}.
(See also \cite{PRD, Les}.)
It gives the density profile of
the system $(\widetilde{\Xi}, \widetilde{\mathrm{P}}_N)$
as
\begin{equation}
\widetilde{\rho}^{\mathrm{mf}}_N(\phi) = \frac{1}{\pi g_s} \tan^{-1}
\left[ \frac{\sqrt{e^{Ng_s} -\big( \cosh \left(\phi/2\right) \big)^2}}
{\cosh \left(\phi/2\right)} \right].
\label{Mdens2}
\end{equation}
Since $\lim_{x\to \infty}\tan^{-1}x = \pi /2$,
in the limit $N \to \infty$
(\ref{Mdens2}) becomes
\begin{equation}
\widetilde{\rho}^{\mathrm{mf}}(\phi) = \frac{1}{2g_s}.
\end{equation}
It is the mean value of the present rigorous result
$\widetilde{\rho}(\phi)$.
As shown in Figure 3, when $g_s$ is much large the density profile
seems to be an equidistant set of peaks, that is,
a lattice structure appears.
In the vicinity of the origin,
\begin{equation}
\widetilde{\rho}(\phi) \sim \widetilde{\rho}(0) \cosh \phi,
\quad\mbox{$|\phi| \ll g_s$},
\end{equation}
with
\begin{equation}
\widetilde{\rho}(0) \equiv
\frac{2e^{-g_s/2}}
{\sqrt{2\pi g_s}}
\to 0, \quad \mbox{as $g_s \to \infty$}.
\end{equation}

In \cite{Tierz2005}, de Haro and Tierz reported a numerical observation
of $N$-dependence of the density profile $\widetilde{\rho}_N^{(1)}(\phi)$
for $(\widetilde{\Xi}, \widetilde{\mathrm{P}}_N)$,
in which the relation (\ref{gs}) between $g_s$ and $N$ was not imposed
and $N$ and $q=e^{-g_s}\in (0,1)$ were treated as free parameters.
They demonstrated that for any finite $N$ with $0<q<1$,
oscillatory behaviors are observed.
They found that, if $q$ is fixed to be less that $1$,
the oscillatory behavior remains even in setting $N$ large.
As claimed by them, however, the relation (\ref{gs}) implies that
$N \to \infty$ limit gives $g_s \to 0$ and $q=e^{-g_s} \to 1$,
and hence the remarkable oscillatory behavior
will be smoothed out in $N \to \infty$.
In the present paper, however, we take the $N \to \infty$ limit
with keeping $0<g_s<\infty$ \textit{i.e.} $0<q<1$.
The obtained determinantal point process
with the correlation kernel $\mathcal{K}_{\infty}$
is a stationary realization of the oscillatory matrix model
of de Haro and Tierz which is constructed uniformly on $\mathbb{R}$.
The present limit $N \to \infty$ with fixed $0<q=e^{-g_s}<1$
has not been able to be rigorously studied so far
because of the lack of suitable $q$-analogous formulas of
the Plancherel-Rotach asymptotics for the Stieltjes-Wigert polynomials
as mentioned in \cite{Tierz2005}.
This problem was solved by Lemmas 1 and 2 in the present paper.
See also \cite{WW06} and \cite{LW13} for other study on the asymptotics
of the Stieltjes-Wigert polynomials.
%
\begin{figure}[htbp]
  \begin{center}
   \includegraphics[clip,width=9.0cm]{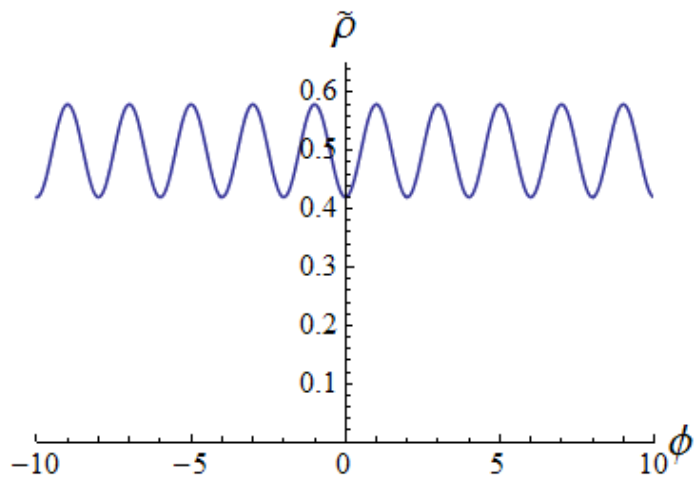}
  \end{center}
  \caption{The density profile $\widetilde{\rho}(\phi)$ with $g_s=1$.
  The period is $2g_s = 2$ and the mean value is $1/2g_s = 0.5$.}
  \label{fig1}
\end{figure}
\begin{figure}[htbp]
  \begin{center}
   \includegraphics[clip,width=9.0cm]{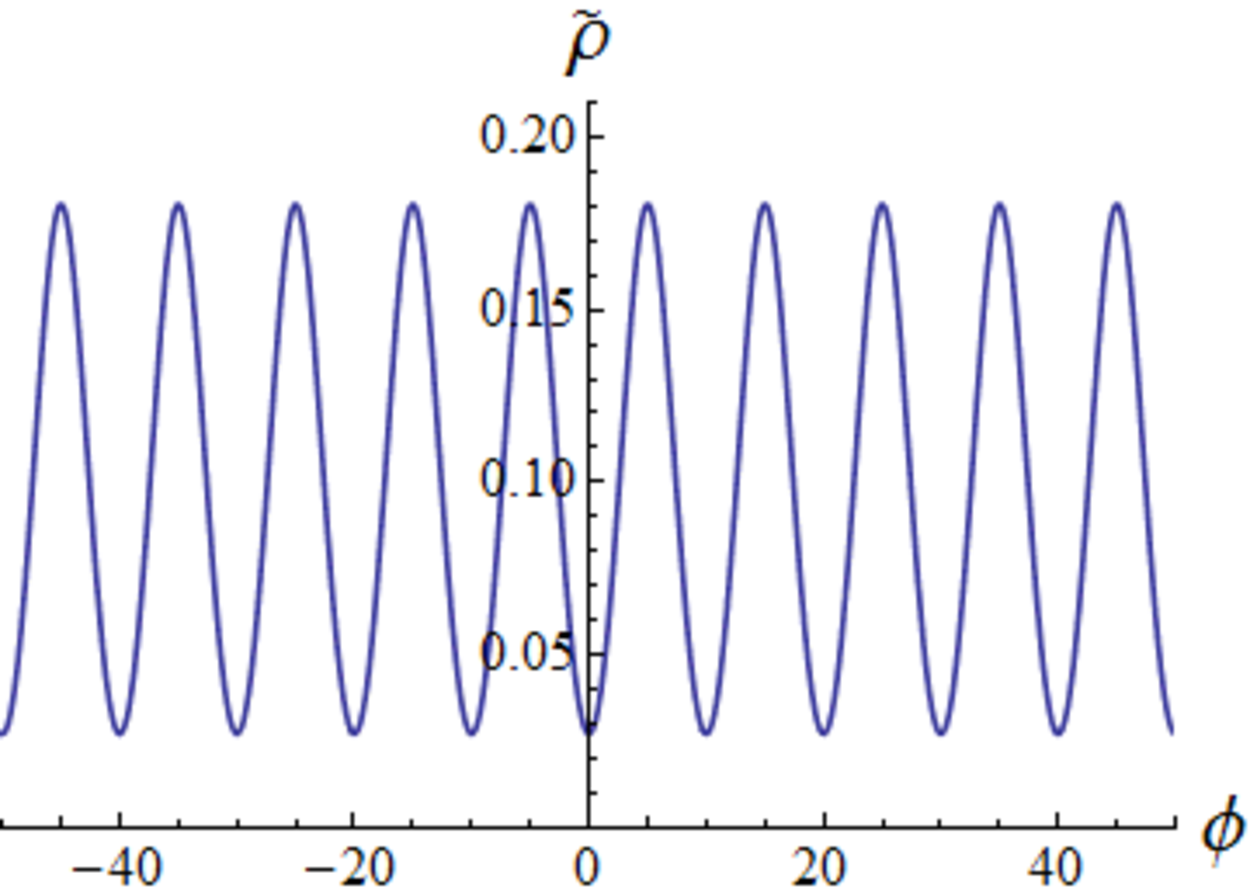}
  \end{center}
  \caption{The density profile $\widetilde{\rho}(\phi)$ with $g_s=5$.
  The period is $2g_s = 10$ and the mean value is $1/2g_s = 0.1$.}
  \label{fig_e}
\end{figure}
\begin{figure}[htbp]
  \begin{center}
   \includegraphics[clip,width=9.0cm]{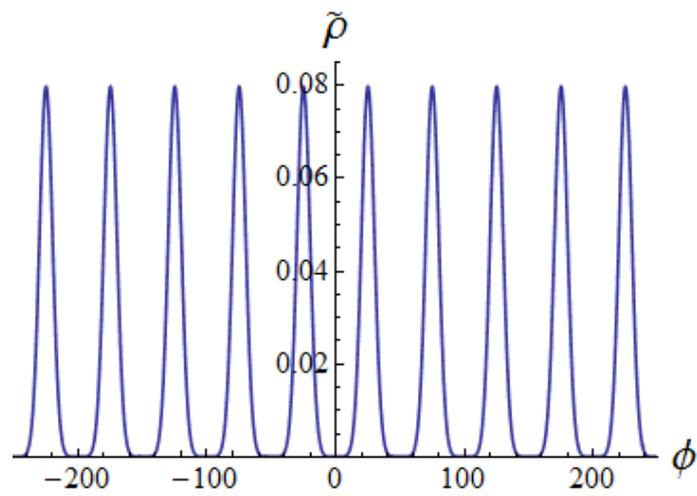}
  \end{center}
  \caption{The density profile $\widetilde{\rho}(\phi)$ with $g_s=25$.
  The period is $2g_s = 50$ and the mean value is $1/2g_s = 0.02$.
  Equidistant peaks imply a lattice
  structure on $\mathbb{R}$. }
  \label{fig2}
\end{figure}

\subsection{Reduction to the sine-kernel}
For consistency with the consideration and observation
by de Haro and Tierz \cite{Tierz2005},
the further limit $g_s \to 0$ ($q\to 1$)
after $N \overset{\tau=1}{\longrightarrow} \infty$
should reduce our oscillatory model to be
the classical one-matrix model.
That is, the determinantal point process with the correlation kernel
$\mathcal{K}_{\infty}$ should converge to that with the sine-kernel.
We consider the sine-kernel with density $1$,
\begin{eqnarray}
K_{\sin} (\phi,\varphi)=
\begin{cases}
\displaystyle{
\frac{\sin(\pi(\phi-\varphi))}{\pi(\phi-\varphi)}},
&\mbox{$\phi,\varphi \in \mathbb{R}$, $\phi \neq \varphi$,}\\
\displaystyle{
1}, &\mbox{$\phi = \varphi \in \mathbb{R}$.}
\end{cases}
\label{sin}
\end{eqnarray}
The following proposition ensures the fact.
\vskip 0.5 cm
\noindent \textbf{Proposition 6} \quad
We have
\begin{equation}
\lim_{g_s \to 0} 2g_s \mathcal{K}_{\infty}
(2g_s \phi,2g_s \varphi) = K_{\sin} (\phi,\varphi).
\label{symmbreak}
\end{equation}

The proof of Proposition 6 is owed to the following
asymptotic expansion for the $q$-exponential function,
which was given as equation (3.13) in \cite{Daalhuis}.
\vskip 0.5cm
\noindent \textbf{Lemma 7}\cite{Daalhuis} \quad
For $|\arg z| \leq 2\pi$,
\begin{eqnarray}
&&\frac{1}{e_q(z)} = \frac{2\sin (\pi \ln z/ \ln q)}
{(z^{-1}q;q)_{\infty}}
\exp \left[ \frac{1}{2} \ln z
- \frac{1}{\ln q} \left\{ -\frac{\pi^2}{3} + \frac{1}{2} (\ln z)^2
\right\}
- \frac{1}{12} \ln q \right. \nonumber \\
&& \quad\quad\quad\quad\quad\quad
\quad\quad\quad\quad\quad\quad\quad\quad\quad\quad\quad
+ \left. \sum_{k=1}^{\infty}
\frac{ \cos \left(2\pi k \ln z/\ln q \right) \exp(2\pi^2 k/\ln q) }
{ k \sinh \left( 2\pi^2 k/\ln q \right) } \right].
\label{e_q}
\end{eqnarray}
\vskip 0.5cm
\noindent
As pointed out in \cite{Daalhuis},
since $1/e_q(z) = (z;q)_{\infty}$,
Lemma 7 gives an expansion formula for
the function $(z;q)_{\infty}(z^{-1}q;q)_{\infty}$.
Hence by Jacobi's triple product identity (\ref{J3})
for $\Theta$, we have
\begin{eqnarray}
\Theta(-z|q) &=& (q^2;q^2)_{\infty}
2\cos \left(\frac{\pi \ln z}{2\ln q} \right)
\exp \left[
- \frac{1}{2\ln q} \left\{ -\frac{\pi^2}{3} 
+ \frac{1}{2} (\ln z)^2 \right\}
+ \frac{1}{12} \ln q \right. \nonumber \\
&\quad& \quad+ \left. \sum_{k=1}^{\infty}
\frac{ \cos \left(\pi k \ln z/\ln q +\pi k \right)
\exp(\pi^2 k/\ln q) }
{ k \sinh \left( \pi^2 k/\ln q \right) } \right], 
\quad \mbox{for $|\arg z| \leq 2\pi$}.
\label{theta_ex}
\end{eqnarray}
Then,
for the theta function used to express the kernel
$\mathcal{K}_{\infty}$ in (\ref{bmkernel})
and (\ref{bmkernel_onep}), we obtain
\begin{eqnarray}
&&\Theta(-e^{\phi \pm g_s/2}|e^{-g_s}) 
= (e^{-2g_s};e^{-2g_s})_{\infty}
2\cos \left(\frac{\pi \phi}{2g_s} \pm \frac{\pi}{4} \right)
\nonumber \\
&&\quad\quad\times
\exp \left[
-\frac{\pi^2}{6g_s} + \frac{\phi^2}{4g_s} \pm \frac{\phi}{4}
- \frac{g_s}{48}
+ \sum_{k=1}^{\infty}
\frac{ \cos \left(\pi k \phi/g_s \pm \pi k/2 - \pi k \right)
\exp(-\pi^2 k/g_s) }
{ k \sinh \left( -\pi^2 k/g_s \right) } \right],
\nonumber \\
&\quad&\quad\quad\quad\quad\quad\quad\quad\quad
\quad\quad\quad\quad\quad\quad\quad\quad\quad\quad
\quad\quad\quad\quad\quad\quad\quad\quad\quad\quad
\quad\quad\quad
\phi \in \mathbb{R}.
\label{theta_exo}
\end{eqnarray}
\vskip 0.5 cm
\noindent
Here, we note
the following asymptotic expansion for the $q$-Pochhammer symbol
(Theorem 2 in \cite{Mcintosh}).
For $q=e^{-\varepsilon}$, $\varepsilon >0$,
\begin{equation}
(q;q)_{\infty} = \sqrt{\frac{2\pi}{\varepsilon}}
e^{-\pi^2/6\varepsilon} (1+ \mathcal{O}(\varepsilon)),
\quad \mbox{as $\varepsilon \to 0$}.
\label{Mcintosh}
\end{equation}
\textit{Proof of Proposition 6}.
Asymptotic expansion of $2g_s \mathcal{K}_{\infty}
(2g_s \phi,2g_s \varphi)$ as $g_s \to 0$ is obtained by
(\ref{theta_exo}) and (\ref{Mcintosh}).
The last term in the exponential in (\ref{theta_exo})
given by an infinite sum
are irrelevant to other terms in $g_s \to 0$.
The contribution to $2g_s \mathcal{K}_{\infty}(2g_s \phi,2g_s \varphi)$
which comes from
$\exp \left(-\pi^2/6g_s \right)$ in (\ref{theta_exo}) is
completely canceled by that from $q$-Pochhammer symbols
in (\ref{bmkernel}), (\ref{bmkernel_onep}), and (\ref{theta_exo})
by the formula (\ref{Mcintosh}).
Therefore, in the limit $g_s \to 0$, only the factor
$\cos \left(\pi \phi/2g_s \pm \pi/4 \right)$ in (\ref{theta_exo})
is relevant, and the proof is completed. \qed
\vskip 0.5cm

In order to demonstrate the difference between the kernel
$\mathcal{K}_{\infty}(\phi, \varphi)$ of the oscillatory
model and $K_{\sin}(\phi, \varphi)$ of the classical model,
we plotted them as functions of $(\phi, \varphi)$ in Figures 3 and 4.
\vskip 0.5cm
\begin{figure}[htbp]
  \begin{center}
   \includegraphics[clip,width=9.0cm]{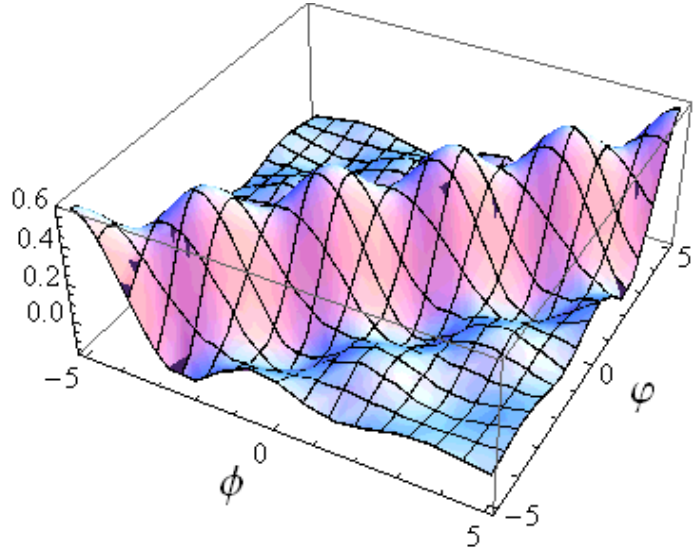}
  \end{center}
  \caption{The correlation kernel $\mathcal{K}_{\infty}(\phi, \varphi)$
   of the oscillatory matrix model as a function of $(\phi, \varphi)$.
   The case with
   $g_s=1$ ($q = e^{-g_s} \simeq 0.37$) is plotted.
   The oscillatory behaviors in the diagonal directions
   show breakdown of the translational invariance.}
  \label{fig3}
\end{figure}
\begin{figure}[htbp]
  \begin{center}
   \includegraphics[clip,width=9.0cm]{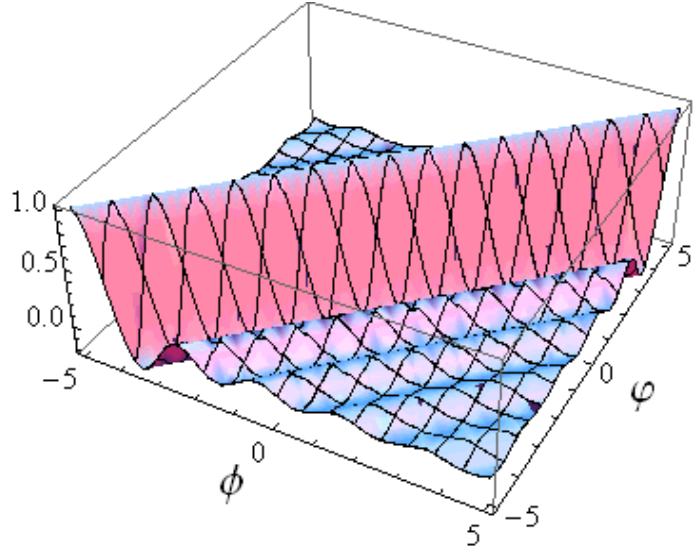}
  \end{center}
  \caption{The sine-kernel $K_{\sin}(\phi, \varphi)$ is plotted
  as a function of $(\phi, \varphi)$.
  The value depends only on $\phi-\varphi$ and the
  translational invariance of the system is shown.}
  \label{fig4}
\end{figure}
\vskip 0.5cm
\newpage
\section{Proofs of Lemma 1 and Proposition 3}
Assume that $0<q<1$.
The following estimate is obtained
from Lemma 3.1 in \cite{IsmailZhang}. Let
\begin{eqnarray}
\frac{(q;q)_{\infty}}{(q;q)_n} = 1+\widetilde{R}(q;n).
\end{eqnarray}
Then
\begin{eqnarray}
|\widetilde{R}(q;n)| \leq \frac{(-q^3;q)_{\infty}}{1-q} q^{n+1}.
\end{eqnarray}
The following lemma improves this estimate.
\vskip 0.5cm
\noindent \textbf{Lemma 8} \quad
Let
\begin{equation}
\frac{(q;q)_{\infty}}{(q;q)_n} = 1-\frac{q^{n+1}}{1-q} + R(q;n).
\label{53}
\end{equation}
Then
\begin{equation}
|R(q;n)|<\frac{(-q;q)_{\infty}}{(1-q)(1-q^2)} q^{2n+2}.
\label{54}
\end{equation}
\vskip 0.5cm
\noindent{\it Proof} \,
From the definition of the $q$-Pochhammer symbol (\ref{qPs}),
we have
\begin{eqnarray}
\frac{(q;q)_{\infty}}{(q;q)_n} =(q^{n+1};q)_{\infty}
= 1 - \frac{q^{n+1}}{1-q} + \sum_{k=2}^{\infty}
\frac{q^{k(k-1)/2}}{(q;q)_k}(-1)^kq^{k(n+1)}.
\end{eqnarray}
Then, by (\ref{53}),
\begin{eqnarray}
R(q;n) &=& \sum_{k=2}^{\infty}
\frac{q^{k(k-1)/2}}{(q;q)_k}(-1)^kq^{k(n+1)}
\nonumber \\
&=& q^{2(n+1)} \sum_{k=0}^{\infty}
\frac{q^{(k+2)(k+1)/2}}{(q;q)_{k+2}}(-1)^kq^{k(n+1)}.
\label{Rqn}
\end{eqnarray}
By taking the absolute value of (\ref{Rqn}), we have
\begin{eqnarray}
|R(q;n)| &<& q^{2(n+1)} \sum_{k=0}^{\infty}
\frac{q^{(k+2)(k+1)/2}}{(q;q)_{k+2}}q^{k(n+1)}
\nonumber \\
&=& q^{2n+2} \sum_{k=0}^{\infty}
\frac{q^{k(k-1)/2}}{(q;q)_{k}} q^k
\frac{q^{2k+1+kn}}{(1-q^{k+1})(1-q^{k+2})}.
\end{eqnarray}
Since for any $n,k \in \mathbb{N}_0$,
$q^{2k+1+kn}<1$ and
$1/(1-q^{k+1})(1-q^{k+2}) \leq 1/(1-q)(1-q^2)$,
we have
\begin{eqnarray}
|R(q;n)| &<& \frac{ q^{2n+2} }{(1-q)(1-q^{2})} \sum_{k=0}^{\infty}
\frac{q^{k(k-1)/2}}{(q;q)_{k}} q^k
\nonumber \\
&=& \frac{(-q;q)_{\infty}}{(1-q)(1-q^{2})} q^{2n+2},
\end{eqnarray}
which proves (\ref{54}). \qed
\vskip 0.5cm
Let
\begin {equation}
S_n(x;q)=\sum_{k=0}^{n}
\frac{q^{k^2}(-x)^k}{(q;q)_{k}(q;q)_{n-k}}
=q^{n^2} (-x)^n\sum_{k=0}^{n}
\frac{q^{k^2 -2kn}(-x)^{-k}}{(q;q)_{k}(q;q)_{n-k}},
\quad  n \in \mathbb{N}_0.
\label{SWpoly}
\end{equation}
The orthonormal Stieltjes-Wigert polynomials (\ref{onSWpoly})
are then given by
\begin{equation}
p_n(x;q)=(-1)^n q^{n/2+1/4}\sqrt{(q;q)_n}S_n(q^{1/2}x;q),
\quad  n \in \mathbb{N}_0.
\label{orth}
\end{equation}
\subsection{Proof of Lemma 1}
We set $x=q^{-n\tau}u$, $u>0$, $\tau \in (0,2)$
and assume (\ref{chi}), (\ref{m}), and (\ref{lambda}).
By definitions (\ref{m}) and (\ref{lambda}),
$m + \lambda = (2-\tau)n$, and we have
\begin{eqnarray}
S_n\left(q^{-n\tau}u;q\right)=
\frac{(-u)^n q^{n^2 (1-\tau)}}{(q;q)_{\infty}^2}
\sum_{k=0}^{n}
\frac{(q;q)_{\infty}^2 q^{k^2}}{(q;q)_{k}(q;q)_{n-k}}
\left(-q^{-m-\lambda} u^{-1}\right)^{k}.
\end{eqnarray}
Here we split the sum $S_n\left(q^{-n\tau}u;q\right)$
into two parts as follows,
\begin{eqnarray}
&&S_n\left(q^{-n\tau}u;q\right) = S^{(1)}_n +S^{(2)}_n,
\label{fo}\\
&&S^{(1)}_n=
\frac{(-u)^n q^{n^2 (1-\tau)}}{(q;q)_{\infty}^2}
\sum_{k=0}^{\lfloor m/2\rfloor} 
\frac{(q;q)_{\infty}^2 q^{k^2}}{(q;q)_{k}(q;q)_{n-k}}
\left(-q^{-m-\lambda} u^{-1}\right)^{k},
\\
&&S^{(2)}_n=
\frac{(-u)^n q^{n^2 (1-\tau)}}{(q;q)_{\infty}^2}
\sum_{k=\lfloor m/2\rfloor +1}^{n} 
\frac{(q;q)_{\infty}^2 q^{k^2}}{(q;q)_{k}(q;q)_{n-k}}
\left(-q^{-m-\lambda} u^{-1}\right)^{k}.
\end{eqnarray}
We rewrite $S_n^{(1)}$ as
\begin{eqnarray}
\hspace{-1.5cm}
&&S_n^{(1)}=\frac{q^{n^2 (1-\tau)
-\lfloor m/2\rfloor (\lfloor m/2\rfloor + \chi(m) + \lambda)}}
{(-u)^{\lfloor m/2 \rfloor -n} (q;q)_{\infty}^2}
\sum_{k=0}^{\lfloor m/2\rfloor} q^{k^2}
\left(-q^{\lambda+\chi(m)} u\right)^{k}
\frac{(q;q)_{\infty}}{(q;q)_{\lfloor m/2\rfloor -k}}
\frac{(q;q)_{\infty}}{(q;q)_{n-\lfloor m/2\rfloor +k}}.
\end{eqnarray}
Applying Lemma 8 yields nine terms as follows,
\begin{eqnarray}
&&S_n^{(1)}=\frac{q^{n^2 (1-\tau)
-\lfloor m/2\rfloor (\lfloor m/2\rfloor + \chi(m) + \lambda)}}
{(-u)^{\lfloor m/2 \rfloor -n} (q;q)_{\infty}^2}
\nonumber \\
&&\qquad\times
\sum_{k=0}^{\lfloor m/2\rfloor} q^{k^2}
\left(-q^{\lambda+\chi(m)} u\right)^{k}
\Bigg( 1 -\frac{q^{n-\lfloor m/2\rfloor +k+1}}{1-q}
-\frac{q^{\lfloor m/2\rfloor -k+1}}{1-q}
\nonumber \\
&&\qquad
+\frac{q^{n+2}}{(1-q)^2}
+R(q;n-\lfloor m/2\rfloor +k)
-\frac{q^{\lfloor m/2\rfloor -k+1}}{1-q}
R(q;n-\lfloor m/2\rfloor +k)
\nonumber \\
&&\qquad
+R(q;\lfloor m/2\rfloor -k)
-\frac{q^{n-\lfloor m/2\rfloor +k+1}}{1-q}
R(q;\lfloor m/2\rfloor -k)
\nonumber \\
&&\qquad\qquad\qquad\qquad\qquad\qquad\qquad\qquad
+R(q;\lfloor m/2\rfloor -k)R(q;n-\lfloor m/2\rfloor +k)
\Bigg).
\end{eqnarray}
Then we write $S_n^{(1)}$ as
\begin{eqnarray}
&&S_n^{(1)}=
\frac{q^{n^2 (1-\tau)
-\lfloor m/2\rfloor (\lfloor m/2\rfloor + \chi(m) + \lambda)}}
{(-u)^{\lfloor m/2 \rfloor -n} (q;q)_{\infty}^2}
\nonumber \\
&&\quad\quad\quad\quad\times
\Bigg\{
\sum_{k=0}^{\infty} q^{k^2} \left(-q^{\lambda+\chi(m)} u\right)^{k}
-\frac{q^{1+n-\lfloor m/2\rfloor}}{1-q}\sum_{k=0}^{\infty} 
q^{k^2 +k} \left(-q^{\lambda+\chi(m)} u\right)^{k}
\nonumber \\
&&\quad\quad\quad\quad\quad\quad\quad\quad
\quad\quad\quad\quad\quad\quad
-\frac{q^{1+\lfloor m/2\rfloor}}{1-q}\sum_{k=0}^{\infty} 
q^{k^2 -k} \left(-q^{\lambda+\chi(m)} u\right)^{k}
+r_1(n) \Bigg\},
\label{s1}
\end{eqnarray}
where $r_1(n)$ consists of nine terms
shown explicitly as (\ref{r1}) in Appendix B.
Similarly, we rewrite $S_n^{(2)}$ as
\begin{eqnarray}
S_n^{(2)}=\frac{q^{n^2 (1-\tau)
-\lfloor m/2\rfloor (\lfloor m/2\rfloor + \chi(m) + \lambda)}}
{(-u)^{\lfloor m/2 \rfloor -n} (q;q)_{\infty}^2}
\sum_{k=1}^{n-\lfloor m/2\rfloor} \frac{(q;q)_{\infty}^2 
q^{k^2} \left(-q^{\lambda+\chi(m)} u\right)^{-k} }
{(q;q)_{\lfloor m/2\rfloor +k}(q;q)_{n-\lfloor m/2\rfloor -k}},
\end{eqnarray}
and through the Lemma 8, we write
\begin{eqnarray}
&&S_n^{(2)}=\frac{q^{n^2 (1-\tau)
-\lfloor m/2\rfloor (\lfloor m/2\rfloor + \chi(m) + \lambda)}}
{(-u)^{\lfloor m/2 \rfloor -n} (q;q)_{\infty}^2}
\nonumber \\
&&\quad\quad \times
\Bigg\{
\sum_{k=1}^{\infty} q^{k^2}
\left(-q^{\lambda+\chi(m)} u\right)^{-k}
-\frac{q^{1+n-\lfloor m/2\rfloor}}{1-q}\sum_{k=1}^{\infty} 
q^{k^2 -k} \left(-q^{\lambda+\chi(m)} u\right)^{-k}
\nonumber \\
&&\quad\quad\quad\quad\quad\quad\quad
\quad\quad\quad\quad\quad\quad\quad
-\frac{q^{1+\lfloor m/2\rfloor}}{1-q}\sum_{k=1}^{\infty} 
q^{k^2 +k} \left(-q^{\lambda+\chi(m)} u\right)^{-k}
+r_2(n) \Bigg\},
\label{s2}
\end{eqnarray}
where $r_2(n)$ consists of nine terms
shown explicitly as (\ref{r2}) in Appendix B.
Thus we have
\begin{eqnarray}
&&S_n^{(1)}+S_n^{(2)}=\frac{q^{n^2 (1-\tau)
-\lfloor m/2\rfloor (\lfloor m/2\rfloor + \chi(m) + \lambda)}}
{(-u)^{\lfloor m/2 \rfloor -n} (q;q)_{\infty}^2}
\nonumber \\
&&\qquad\times
\Bigg\{
\sum_{k=-\infty}^{\infty} q^{k^2}
\left(-q^{\lambda+\chi(m)} u\right)^{k}
-\frac{q^{1+n-\lfloor m/2\rfloor}}{1-q}\sum_{k=-\infty}^{\infty} 
q^{k^2 +k} \left(-q^{\lambda+\chi(m)} u\right)^{k}
\nonumber \\
&&\qquad\qquad\qquad\qquad
-\frac{q^{1+\lfloor m/2\rfloor}}{1-q}\sum_{k=-\infty}^{\infty} 
q^{k^2 -k} \left(-q^{\lambda+\chi(m)} u\right)^{k}
+r_1(n)+r_2(n) \Bigg\}.
\label{s1s2}
\end{eqnarray}
Then infinite sums in the parenthesis in (\ref{s1s2})
can be expressed by using the theta functions (\ref{thetafnc}).
Therefore, by (\ref{fo}), we obtain
\begin{eqnarray}
&&\hspace{-0.8cm}S_n \left(q^{-n\tau}u;q\right)=\frac{q^{n^2 (1-\tau)
-\lfloor m/2\rfloor (\lfloor m/2\rfloor + \chi(m) + \lambda)}}
{(-u)^{\lfloor m/2 \rfloor -n} (q;q)_{\infty}^2}
\Bigg\{ \Theta\left(-q^{\lambda+\chi(m)} u\Big|q\right)
\nonumber\\
&&\hspace{-0.8cm}
-\frac{q^{1+n-\lfloor m/2\rfloor}}{1-q}
\Theta\left(-q^{\lambda+\chi(m)+1} u\Big|q\right)
-\frac{q^{1+\lfloor m/2\rfloor}}{1-q}
\Theta\left(-q^{\lambda+\chi(m)-1} u\Big|q\right)
+r_1(n)+r_2(n) \Bigg\}.
\label{fom}
\end{eqnarray}
It is shown in Appendix B that
the terms $r_1(n) + r_2(n)$
can be evaluated as
\begin{equation}
r_1(n) + r_2(n) =
\mathcal{O}\left(q^{\tau n +2(1-\tau)n \I_{[1,2)}(\tau)}\right).
\label{order}
\end{equation}
Since
\begin{equation}
\frac{\tau n}{2} \leq n-\lfloor m/2\rfloor < \frac{\tau n}{2} +1
\end{equation}
and
\begin{equation}
\frac{(2-\tau)n}{2} -1< \lfloor m/2\rfloor \leq \frac{(2-\tau)n}{2},
\label{b4}
\end{equation}
we have to care about
the order of $ \tau n/2$, $(2-\tau)n/2$, $\tau n$, and $(2-\tau)n$.
We can readily see that in the case of $2/3 < \tau < 4/3$,
equation (\ref{fom}) holds with (\ref{order}) as it stands.
In the case of $0 < \tau \leq 2/3$, the last two relevant terms
in the parenthesis in (\ref{fom})
can be replaced by
$-\frac{q^{1+n-\lfloor m/2\rfloor}}{1-q}
\Theta\left(-q^{\lambda+\chi(m)+1} u\big|q\right)$ and the estimate
${\cal O}\big(q^{\tau n +2(1-\tau)n \I_{[1,2)}(\tau)}\big)$
for the irrelevant terms (\ref{order}) \hs
by \hs ${\cal O}\big(q^{\tau n}\big)$. \hs
In \hs the \hs case \hs of \hs $4/3 \leq \tau < 2$, \hs
the \hs former \hs can \hs be \hs replaced \hs by\\
$-\frac{q^{1+\lfloor m/2\rfloor}}{1-q}
\Theta\big(-q^{\lambda+\chi(m)-1} u\big|q\big)$
and the latter
${\cal O}\big(q^{\tau n +2(1-\tau)n \I_{[1,2)}(\tau)}\big)$
by ${\cal O}\big(q^{(2-\tau) n}\big)$.
Hence, for the orthonormal Stieltjes-Wigert polynomials (\ref{orth}),
we have
\begin{eqnarray}
&&p_n \left(q^{-n\tau}u;q\right)=(-1)^n q^{n/2 +1/4}\sqrt{(q;q)_n}
S_n\left(q^{-n\tau}q^{1/2}u;q\right)
\nonumber \\
&&
=\frac{(-1)^n q^{n/2+1/4 +n^2 (1-\tau)
-\lfloor m/2\rfloor (\lfloor m/2\rfloor + \chi(m) + \lambda)}
\sqrt{(q;q)_n}}
{(-q^{1/2}u)^{\lfloor m/2 \rfloor -n} (q;q)_{\infty}^2}
\nonumber\\
&&\quad\times
\left\{ \Theta\Big(-q^{\lambda+\chi(m)+1/2} u\Big|q\Big)
-\frac{q^{1+n-\lfloor m/2\rfloor}}{1-q}
\Theta\Big(-q^{\lambda+\chi(m)+3/2} u\Big|q\Big) \I_{(0,4/3)}(\tau)
\right.
\nonumber\\
&&\quad\quad\quad
\left.
-\frac{q^{1+\lfloor m/2\rfloor}}{1-q}
\Theta\Big(-q^{\lambda+\chi(m)-1/2} u\Big|q\Big)\I_{(2/3,2)}(\tau)
+{\cal O}\left(q^{\tau n +2(1-\tau)n \I_{[1,2)}(\tau)}\right) \right\}.
\end{eqnarray}
Further, by noting the functional equation (\ref{fnceqT})
and $\lfloor m/2\rfloor = m/2 -\chi(m)/2$ ,
(\ref{onSWayex}) is obtained. It completes the proof. \qed
\vskip 0.5cm
\subsection{Proof of Proposition 3}
By Lemma 2 and the definition of $N \overset{\tau}{\to} \infty$,
we have
\begin{eqnarray}
\hspace{-0.5cm}
&&\hspace{-0.5cm}p_{N-1} \left(q^{-\lceil \tau N \rceil 
-\chi \left(\lceil \tau N \rceil \right)}v;q \right)
\sqrt{w\left(q^{-\lceil \tau N \rceil 
-\chi \left(\lceil \tau N \rceil \right)}v;q \right)}
\nonumber \\
&&\hspace{-0.5cm}\quad
=\frac{ \sqrt{(q;q)_{N-1}} }{(q;q)_{\infty}^2}(-1)^{N-1}
q^{N/2-1/4+\lceil \tau N \rceil /4
+ \chi \left(\lceil \tau N \rceil \right)/4} \sqrt{w(v;q)}
\Bigg\{ \Theta\Big(-q^{1/2} v\Big|q\Big)
\nonumber \\
&&\hspace{-0.5cm}\quad
+\frac{q^{N - \lceil \tau N \rceil /2
- \chi \left(\lceil \tau N \rceil \right)/2}}
{1-q}
\left( \frac{q^{1/2-N+ \lceil \tau N\rceil
+\chi \left(\lceil \tau N \rceil \right)}}
{v} \I_{(0,4/3)}(\tau)- \I_{(2/3,2)}(\tau) \right)
\Theta\Big(-q^{-1/2} v\Big|q\Big)
\nonumber\\
&&\hspace{-0.5cm}\quad\quad\quad\quad\quad\quad\quad\quad\quad
\quad\quad\quad\quad\quad\quad\quad\quad\quad
+{\cal O}\left(q^{\tau N +2(1-\tau)N \I_{[1,2)}(\tau)}\right) \Bigg\},
\quad
\mbox{as $N \overset{\tau}{\to} \infty$.}
\label{vvv}
\end{eqnarray}
In the parenthesis in (\ref{vvv}),
there are two terms expressed by using $\Theta$,
in which the first term is the leading term and the second one includes
indicators $\I_{(0,4/3)}(\tau)$ and $\I_{(2/3,2)}(\tau)$,
and an irrelevant term of order
${\cal O}\big(q^{\tau N +2(1-\tau)N \I_{[1,2)}(\tau)}\big)$.
We put (\ref{vvv}) and the similar estimate of
$p_N \left(q^{-\lceil \tau N \rceil 
-\chi \left(\lceil \tau N \rceil \right)}u;q \right)$
into the product
\begin{eqnarray}
&&p_N \left(q^{-\lceil \tau N \rceil 
-\chi \left(\lceil \tau N \rceil \right)}u;q \right)
\sqrt{w\left(q^{-\lceil \tau N \rceil 
-\chi \left(\lceil \tau N \rceil \right)}u;q \right)}
\nonumber \\
&&\qquad\qquad\times
p_{N-1} \left(q^{-\lceil \tau N \rceil 
-\chi \left(\lceil \tau N \rceil \right)}v;q \right)
\sqrt{w\left(q^{-\lceil \tau N \rceil 
-\chi \left(\lceil \tau N \rceil \right)}v;q \right)}.
\end{eqnarray}
We find the product of the leading terms expressed by $\Theta$
is symmetric in exchanging $u$ and $v$,
and the product of the second terms
expressed by $\Theta$ with indicators
becomes irrelevant since it is in order
$q^{\tau N +2(1-\tau)N \I_{[1,2)}(\tau)}$.
Then, when we consider the Christoffel-Darboux kernel (\ref{cd}),
the leading terms are canceled out
and the cross terms of the first terms with $\Theta$
and the second terms with $\Theta$ and indicators become relevant.
Moreover, we find that the cross terms which possess the indicator
$\I_{(0,4/3)}(\tau)$ are also completely canceled.
This is the reason why we assume $2/3 < \tau <2$ in this proposition.
Therefore, if we take the $N \overset{\tau}{\to} \infty$ limit,
we obtain the Jacobi-theta kernel. Then the proof is completed. \qed
\section*{Acknowledgments}
MK is supported in part by
the grant-in-aid for Scientific Research (C)
(Grant No.21540397 and No.26400405) of the Japan Society for
the Promotion of Science.
\appendix
\renewcommand{\theequation}{A.\arabic{equation}}
\setcounter{equation}{0}
\def\thesection{Appendix \Alph{section}:}
\section{Expressions of $\mathcal{K}_{\infty}$ using
Jacobi's theta functions and Gosper's
$q$-trigonometric functions}
Let $z=e^{2i\zeta}$, $\zeta \in \mathbb{C}$,
and $q=e^{\pi i \omega}$, $\Im \omega >0$.
The theta function defined by (\ref{thetafnc})
is written by Jacobi's theta function $\vartheta_3$ \cite{WW} as
\begin{equation}
\Theta(z|q) = \vartheta_3(\zeta|\omega)
= \sum _{k=-\infty}^{\infty} e^{k^2 \pi i \omega  + 2ki\zeta }.
\end{equation}
Theta functions used to express $\mathcal{K}_{\infty}$ in
(\ref{bmkernel}) are written by
\begin{equation}
\Theta ( -e^{\phi \pm g_s/2} | e^{-g_s} )=
\sum_{k=-\infty}^{\infty}(-1)^k e^{-k^2 g_s }
e^{k(\phi \pm g_s/2)}
= \vartheta_4 \left( \frac{1}{2i}
\left( \phi \pm \frac{g_s}{2} \right) 
\bigg| \frac{ig_s}{\pi} \right),
\quad\mbox{$\phi \in \mathbb{R}$},
\label{apth}
\end{equation}
where $\vartheta_4$ is defined by
\begin{eqnarray}
\vartheta_4(\zeta|\omega) =
\vartheta_3 \left(\zeta-\frac{\pi}{2} {\Big |} \omega \right)
= \sum_{k=-\infty}^{\infty} (-1)^k e^{k^2 \pi i \omega +2ki\zeta}.
\end{eqnarray}
The quasi-periodicity (\ref{pseudop}) comes from the equality
\begin{equation}
\vartheta_4(\zeta+\omega \pi|\omega) = -e^{-i\pi \omega -2i\zeta}
\vartheta_4(\zeta|\omega).
\label{pseu_T}
\end{equation}
Jacobi's imaginary transformation \cite{WW} for $\vartheta_4$
reads
\begin{equation}
\vartheta_4 (\zeta|\omega)=\frac{1}{\sqrt{-i\omega}}
e^{\zeta^2/\pi i \omega}
\vartheta_2\left( \frac{\zeta}{\omega} {\bigg |}
-\frac{1}{\omega} \right),
\label{jitr}
\end{equation}
where $\vartheta_2$ is defined by
\begin{equation}
\vartheta_2(\zeta|\omega)=e^{i\zeta+\pi i \omega/4}
\vartheta_4(\zeta + \pi/2 + \pi \omega/2|\omega) =
\sum_{k=-\infty}^{\infty} e^{(k+1/2)^2\pi i \omega + (2k+1) i\zeta}.
\end{equation}
Theta functions $\vartheta_2$, $\vartheta_3$, and $\vartheta_4$
are all even functions of $\zeta$.

Gosper's $q$-trigonometric functions are defined as \cite{Gosper}
\begin{eqnarray}
\sin_q (\pi z) &=& q^{(z-1/2)^2} (q^{2z};q^2)_{\infty}
(q^{-2z+2};q^2)_{\infty} (q;q^2)_{\infty}^{-2},
\label{sinq} \\
\cos_q (\pi z) &=&  \sin_q (\pi(z+1/2))
\nonumber \\
&=& q^{z^2} (q^{2z+1};q^2)_{\infty}
(q^{-2z+1};q^2)_{\infty} (q;q^2)_{\infty}^{-2}.
\label{cosq}
\end{eqnarray}
It is easy to find the following properties,
\begin{equation}
\sin_q (-z) = -\sin_q (z), \quad \cos_q (-z) = \cos_q (z),
\end{equation}
\begin{equation}
\sin_q (z+\pi) = -\sin_q (z), \quad \cos_q (z+\pi) = -\cos_q (z).
\end{equation}
By using the product form of $\vartheta_4$
(put $q=e^{\pi i \omega}$ and $z=-e^{2i \zeta}$ in (\ref{J3})),
we have the equality
\begin{equation}
\vartheta_4 (\zeta | \omega) = e^{\zeta^2/\pi i \omega}
(e^{\pi i \omega};e^{\pi i \omega})_{\infty}
(e^{\pi i \omega};e^{2\pi i \omega})_{\infty}
\cos_{e^{\pi i \omega}} \left( \frac{\zeta}{\omega} \right).
\end{equation}
By (\ref{jitr}) we also find
\begin{equation}
\frac{1}{\sqrt{-i\omega}}
\vartheta_2\left( \frac{\zeta}{\omega} {\bigg |} -\frac{1}{\omega} \right)
=(e^{\pi i \omega};e^{\pi i \omega})_{\infty}
(e^{\pi i \omega};e^{2\pi i \omega})_{\infty}
\cos_{e^{\pi i \omega}} \left( \frac{\zeta}{\omega} \right).
\label{th2Gos}
\end{equation}

Then $\mathcal{K}_{\infty}$ is rewritten in terms of $\vartheta_4$,
$\vartheta_2$, and Gosper's $q$-trigonometric functions
as follows,
\begin{eqnarray}
&&\hspace{-0.8cm}\mathcal{K}_{\infty}(\phi,\varphi)
=\frac{1}{\sqrt{2\pi g_s}} \frac{e^{-(\phi^2+\varphi^2)/4g_s} }
{(e^{-g_s};e^{-g_s})_{\infty}^{3} }
\nonumber \\
&&\hspace{-0.8cm}\times \frac{\displaystyle{\vartheta_4 \left( \frac{1}{2i} 
\left( \phi - \frac{g_s}{2} \right) \bigg| \frac{ig_s}{\pi} \right)
\vartheta_4 \left( \frac{1}{2i} \left( \varphi + \frac{g_s}{2} \right)
 \bigg| \frac{ig_s}{\pi} \right)
-\vartheta_4 \left( \frac{1}{2i} \left( \varphi - \frac{g_s}{2} \right)
 \bigg| \frac{ig_s}{\pi} \right)
\vartheta_4 \left( \frac{1}{2i} \left( \phi + \frac{g_s}{2} \right)
 \bigg| \frac{ig_s}{\pi} \right) }}
{\displaystyle{2\sinh \frac{\phi-\varphi}{2}}}
\nonumber
\\
&&\hspace{-0.8cm}=\frac{1}{g_s}\sqrt{\frac{\pi}{2 g_s}} \frac{e^{g_s/8} }
{(e^{-g_s};e^{-g_s})_{\infty}^{3} }
\nonumber \\
&&\hspace{-0.8cm}\times \frac{\displaystyle{ e^{-(\phi-\varphi)/4} \vartheta_2 \left(
\frac{\pi \phi}{2g_s} - \frac{\pi}{4} \bigg| \frac{i\pi}{g_s} \right)
\vartheta_2 \left(\frac{\pi \varphi}{2g_s} + \frac{\pi}{4}
\bigg| \frac{i\pi}{g_s} \right)
-e^{(\phi-\varphi)/4} \vartheta_2 \left(
\frac{\pi \varphi}{2g_s} - \frac{\pi}{4} \bigg| \frac{i\pi}{g_s} \right)
\vartheta_2 \left(\frac{\pi \phi}{2g_s} + \frac{\pi}{4}
\bigg| \frac{i\pi}{g_s} \right)
}}
{\displaystyle{2\sinh \frac{\phi-\varphi}{2}}}
\nonumber \\
&&\hspace{-0.8cm}=\frac{e^{g_s/8}}{\sqrt{2\pi g_s}}
\frac{(e^{-g_s};e^{-2g_s})_{\infty}}{(e^{-2g_s};e^{-2g_s})_{\infty}}
\nonumber \\
&&\hspace{-0.8cm}\times \frac{\displaystyle{ e^{-(\phi-\varphi)/4}
\sin_{e^{-g_s}} \left( \frac{\pi \phi}{2g_s} +\frac{\pi}{4} \right)
\cos_{e^{-g_s}} \left( \frac{\pi \varphi}{2g_s} +\frac{\pi}{4} \right)
-
e^{(\phi-\varphi)/4}
\sin_{e^{-g_s}} \left( \frac{\pi \varphi}{2g_s} +\frac{\pi}{4} \right)
\cos_{e^{-g_s}} \left( \frac{\pi \phi}{2g_s} +\frac{\pi}{4} \right)
}}
{\displaystyle{2\sinh \frac{\phi-\varphi}{2}}},
\nonumber
\end{eqnarray}
$(\phi, \varphi) \in \mathbb{R}^2$.
\renewcommand{\theequation}{B.\arabic{equation}}
\setcounter{equation}{0}
\def\thesection{Appendix \Alph{section}: Proof of (\ref{order})}
\section{ }
The last term in (\ref{s1}) is given by
\begin{equation}
r_1(n) = \sum_{j=1}^9 r_{1j}(n),
\label{r1}
\end{equation}
where
\begin{equation}
r_{11}(n)=-\sum_{k=\lfloor m/2\rfloor+1}^{\infty} q^{k^2} 
\left(-q^{\lambda+\chi(m)} u\right)^{k}, \nonumber
\end{equation}
\begin{equation}
r_{12}(n)=\frac{q^{1+n-\lfloor m/2\rfloor}}{1-q}\sum_{k=\lfloor m/2\rfloor+1}^{\infty} 
q^{k^2 +k} \left(-q^{\lambda+\chi(m)} u\right)^{k}, \nonumber
\end{equation}
\begin{equation}
r_{13}(n)=\frac{q^{1+\lfloor m/2\rfloor}}{1-q}\sum_{k=\lfloor m/2\rfloor+1}^{\infty} 
q^{k^2 -k} \left(-q^{\lambda+\chi(m)} u\right)^{k}, \nonumber
\end{equation}

\begin{equation}
r_{14}(n)=\frac{q^{2+n}}{(1-q)^2}\sum_{k=0}^{\lfloor m/2\rfloor} q^{k^2}
\left(-q^{\lambda+\chi(m)} u\right)^{k}, \nonumber
\end{equation}
and
\begin{equation}
r_{15}(n)=\sum_{k=0}^{\lfloor m/2\rfloor} q^{k^2} 
\left(-q^{\lambda+\chi(m)} u\right)^{k} R(q;n-\lfloor m/2\rfloor +k), \nonumber
\end{equation}

\begin{equation}
r_{16}(n)=-\frac{q^{1+\lfloor m/2\rfloor}}{1-q}
\sum_{k=0}^{\lfloor m/2\rfloor} q^{k^2 -k} 
\left(-q^{\lambda+\chi(m)} u\right)^{k} R(q;n-\lfloor m/2\rfloor +k), \nonumber
\end{equation}
\begin{equation}
r_{17}(n)=\sum_{k=0}^{\lfloor m/2\rfloor} q^{k^2} 
\left(-q^{\lambda+\chi(m)} u\right)^{k} R(q;\lfloor m/2\rfloor -k), \nonumber
\end{equation}
\begin{equation}
r_{18}(n)=-\frac{q^{1+n-\lfloor m/2\rfloor}}{1-q}
\sum_{k=0}^{\lfloor m/2\rfloor} q^{k^2 +k} 
\left(-q^{\lambda+\chi(m)} u\right)^{k} R(q;\lfloor m/2\rfloor -k), \nonumber
\end{equation}
\begin{equation}
r_{19}(n)=\sum_{k=0}^{\lfloor m/2\rfloor} q^{k^2} 
\left(-q^{\lambda+\chi(m)} u\right)^{k} R(q;\lfloor m/2\rfloor -k)
R(q;n-\lfloor m/2\rfloor +k). \nonumber
\end{equation}
Noting $q\in(0,1)$, we have the following inequalities,
\begin{eqnarray}
|r_{11}(n)|&<&\sum_{k=\lfloor m/2\rfloor+1}^{\infty} q^{k^2} 
\left(q^{\lambda+\chi(m)} u\right)^{k} \nonumber \\
&=&q^{\lfloor m/2\rfloor^2 +2\lfloor m/2\rfloor}
\left(q^{\lambda+\chi(m)} u\right)^{\lfloor m/2\rfloor+1}
\sum_{k=0}^{\infty} q^{k^2+2k+1} q^{2\lfloor m/2\rfloor k}
\left(q^{\lambda+\chi(m)} u\right)^{k} \nonumber \\
&<&q^{\lfloor m/2\rfloor^2 +2\lfloor m/2\rfloor}
\left(q^{\lambda+\chi(m)} u\right)^{\lfloor m/2\rfloor+1}
\sum_{k=0}^{\infty} q^{k^2-2k} 
\left(q^{\lambda+\chi(m)} u\right)^{k} \nonumber \\
&<&\frac{q^{\lfloor m/2\rfloor^2 +2\lfloor m/2\rfloor}}
{1-q}\left(q^{\lambda+\chi(m)} u\right)^{\lfloor m/2\rfloor+1}
\sum_{k=0}^{\infty} q^{k^2-2k} \left(q^{\lambda+\chi(m)} u\right)^{k}, \nonumber
\end{eqnarray}
and
\begin{equation}
|r_{12}(n)|<\frac{q^{\lfloor m/2\rfloor^2 +2\lfloor m/2\rfloor +1+n}}{1-q}
\left(q^{\lambda+\chi(m)} u\right)^{\lfloor m/2\rfloor+1}
\sum_{k=0}^{\infty} q^{k^2-2k} \left(q^{\lambda+\chi(m)} u\right)^{k}, \nonumber
\end{equation}
\begin{equation}
|r_{13}(n)|<\frac{q^{\lfloor m/2\rfloor^2 +2\lfloor m/2\rfloor +1}}{1-q}
\left(q^{\lambda+\chi(m)} u\right)^{\lfloor m/2\rfloor+1}
\sum_{k=0}^{\infty} q^{k^2-2k} \left(q^{\lambda+\chi(m)} u\right)^{k}, \nonumber
\end{equation}
\begin{equation}
|r_{14}(n)|<\frac{(-q;q)_{\infty}^2}{(1-q)^2 (1-q^2)^2} q^{n}
\sum_{k=0}^{\infty} q^{k^2-2k} \left(q^{\lambda+\chi(m)} u\right)^{k}. \nonumber
\end{equation}
For the others, applying Lemma 8 yields
\begin{eqnarray}
|r_{15}(n)|&<&\frac{(-q;q)_{\infty}}{(1-q)(1-q^2)} q^{2+2(n-\lfloor m/2\rfloor)}
\sum_{k=0}^{\infty} q^{k^2} \left(q^{\lambda+\chi(m)} u\right)^{k} \nonumber \\
&<&\frac{(-q;q)_{\infty}^2}{(1-q)^2 (1-q^2)^2} q^{2(n-\lfloor m/2\rfloor)}
\sum_{k=0}^{\infty} q^{k^2-2k} \left(q^{\lambda+\chi(m)} u\right)^{k}, \nonumber
\end{eqnarray}
and
\begin{eqnarray}
|r_{16}(n)|
&<&\frac{(-q;q)_{\infty}^2}{(1-q)^2 (1-q^2)^2} q^{2n-\lfloor m/2\rfloor}
\sum_{k=0}^{\infty} q^{k^2-2k} \left(q^{\lambda+\chi(m)} u\right)^{k}, \nonumber
\end{eqnarray}
\begin{eqnarray}
|r_{17}(n)|
&<&\frac{(-q;q)_{\infty}^2}{(1-q)^2 (1-q^2)^2} q^{2\lfloor m/2\rfloor}
\sum_{k=0}^{\infty} q^{k^2-2k} \left(q^{\lambda+\chi(m)} u\right)^{k}, \nonumber
\end{eqnarray}
\begin{eqnarray}
|r_{18}(n)|
&<&\frac{(-q;q)_{\infty}^2}{(1-q)^2 (1-q^2)^2} q^{n+\lfloor m/2\rfloor}
\sum_{k=0}^{\infty} q^{k^2-2k} \left(q^{\lambda+\chi(m)} u\right)^{k}, \nonumber
\end{eqnarray}
\begin{eqnarray}
|r_{19}(n)|
&<&\frac{(-q;q)_{\infty}^2}{(1-q)^2 (1-q^2)^2} q^{2n}
\sum_{k=0}^{\infty} q^{k^2-2k} \left(q^{\lambda+\chi(m)} u\right)^{k}. \nonumber
\end{eqnarray}
On the other hand, the last term in (\ref{s2}) is given by
\begin{equation}
r_2(n)= \sum_{j=1}^9 r_{2j}(n),
\label{r2}
\end{equation}
where
\begin{equation}
r_{21}(n)=-\sum_{k=n-\lfloor m/2\rfloor+1}^{\infty} q^{k^2} 
\left(-q^{\lambda+\chi(m)} u\right)^{-k} \nonumber,
\end{equation}
\begin{equation}
r_{22}(n)=\frac{q^{1+n-\lfloor m/2\rfloor}}{1-q}\sum_{k=n-\lfloor m/2\rfloor+1}^{\infty} 
q^{k^2 -k} \left(-q^{\lambda+\chi(m)} u\right)^{-k} \nonumber,
\end{equation}
\begin{equation}
r_{23}(n)=\frac{q^{1+\lfloor m/2\rfloor}}{1-q}\sum_{k=n-\lfloor m/2\rfloor+1}^{\infty}
 q^{k^2 +k} \left(-q^{\lambda+\chi(m)} u\right)^{-k} \nonumber,
\end{equation}

\begin{equation}
r_{24}(n)=\frac{q^{2+n}}{(1-q)^2}\sum_{k=1}^{n-\lfloor m/2\rfloor} q^{k^2}
\left(-q^{\lambda+\chi(m)} u\right)^{-k} \nonumber,
\end{equation}
and
\begin{equation}
r_{25}(n)=\sum_{k=1}^{n-\lfloor m/2\rfloor} q^{k^2} 
\left(-q^{\lambda+\chi(m)} u\right)^{-k} R(q;n-\lfloor m/2\rfloor -k) \nonumber,
\end{equation}

\begin{equation}
r_{26}(n)=-\frac{q^{1+\lfloor m/2\rfloor}}{1-q}
\sum_{k=1}^{n-\lfloor m/2\rfloor} q^{k^2 +k} 
\left(-q^{\lambda+\chi(m)} u\right)^{-k} R(q;n-\lfloor m/2\rfloor -k) \nonumber,
\end{equation}
\begin{equation}
r_{27}(n)=\sum_{k=1}^{n-\lfloor m/2\rfloor} q^{k^2} 
\left(-q^{\lambda+\chi(m)} u\right)^{-k} R(q;\lfloor m/2\rfloor +k) \nonumber,
\end{equation}
\begin{equation}
r_{28}(n)=-\frac{q^{1+n-\lfloor m/2\rfloor}}{1-q}
\sum_{k=1}^{n-\lfloor m/2\rfloor} q^{k^2 -k} 
\left(-q^{\lambda+\chi(m)} u\right)^{-k} R(q;\lfloor m/2\rfloor +k) \nonumber,
\end{equation}
\begin{equation}
r_{29}(n)=\sum_{k=1}^{n-\lfloor m/2\rfloor} q^{k^2} 
\left(-q^{\lambda+\chi(m)} u\right)^{-k} R(q;\lfloor m/2\rfloor +k)
R(q;n-\lfloor m/2\rfloor -k) \nonumber.
\end{equation}
Similarly, by noting $q \in (0,1)$, we have
\begin{eqnarray}
|r_{21}(n)| < \frac{q^{(n-\lfloor m/2\rfloor)^2}}{1-q}
\left(q^{\lambda+\chi(m)} u\right)^{-(n-\lfloor m/2\rfloor)}
\sum_{k=1}^{\infty} q^{k^2-2k} \left(q^{\lambda+\chi(m)} u\right)^{-k}, \nonumber
\end{eqnarray}
\begin{equation}
|r_{22}(n)|<\frac{q^{(n-\lfloor m/2\rfloor)^2 +1}}{1-q}
\left(q^{\lambda+\chi(m)} u\right)^{-(n-\lfloor m/2\rfloor)}
\sum_{k=1}^{\infty} q^{k^2-2k} \left(q^{\lambda+\chi(m)} u\right)^{-k}, \nonumber
\end{equation}
\begin{equation}
|r_{23}(n)|<\frac{q^{(n-\lfloor m/2\rfloor)^2 +1+n}}{1-q}
\left(q^{\lambda+\chi(m)} u\right)^{-(n-\lfloor m/2\rfloor)}
\sum_{k=1}^{\infty} q^{k^2-2k} \left(q^{\lambda+\chi(m)} u\right)^{-k}, \nonumber
\end{equation}
\begin{equation}
|r_{24}(n)|
<\frac{(-q;q)_{\infty}^2}{(1-q)^2 (1-q^2)^2} q^{n}
\sum_{k=1}^{\infty} q^{k^2-2k} \left(q^{\lambda+\chi(m)} u\right)^{-k}. \nonumber
\end{equation}
For the others, applying Lemma 8 yields
\begin{eqnarray}
|r_{25}(n)|
&<&\frac{(-q;q)_{\infty}^2}{(1-q)^2 (1-q^2)^2} q^{2(n-\lfloor m/2\rfloor)}
\sum_{k=1}^{\infty} q^{k^2-2k} \left(q^{\lambda+\chi(m)} u\right)^{-k}, \nonumber
\end{eqnarray}
\begin{eqnarray}
|r_{26}(n)|
&<&\frac{(-q;q)_{\infty}^2}{(1-q)^2 (1-q^2)^2} q^{2n-\lfloor m/2\rfloor}
\sum_{k=1}^{\infty} q^{k^2-2k} \left(q^{\lambda+\chi(m)} u\right)^{-k}, \nonumber
\end{eqnarray}
\begin{eqnarray}
|r_{27}(n)|
&<&\frac{(-q;q)_{\infty}^2}{(1-q)^2 (1-q^2)^2} q^{2\lfloor m/2\rfloor}
\sum_{k=1}^{\infty} q^{k^2-2k} \left(q^{\lambda+\chi(m)} u\right)^{-k}, \nonumber
\end{eqnarray}
\begin{eqnarray}
|r_{28}(n)|
&<&\frac{(-q;q)_{\infty}^2}{(1-q)^2 (1-q^2)^2} q^{n+\lfloor m/2\rfloor}
\sum_{k=1}^{\infty} q^{k^2-2k} \left(q^{\lambda+\chi(m)} u\right)^{-k}, \nonumber
\end{eqnarray}
\begin{eqnarray}
|r_{29}(n)|
&<&\frac{(-q;q)_{\infty}^2}{(1-q)^2 (1-q^2)^2} q^{2n}
\sum_{k=1}^{\infty} q^{k^2-2k} \left(q^{\lambda+\chi(m)} u\right)^{-k}. \nonumber
\end{eqnarray}
Therefore we have
\begin{eqnarray}
|r_1(n) + r_2(n)|
&\leq& \sum_{j=1}^9 |r_{1j}(n)| + \sum_{j=1}^9 |r_{2j}(n)|
\nonumber \\
&<&
\left(1+q^{1+n}+q\right)
\frac{q^{\lfloor m/2\rfloor^2 +2\lfloor m/2\rfloor}}{1-q}
\left(q^{\lambda+\chi(m)} u\right)^{\lfloor m/2\rfloor+1}
\sum_{k=0}^{\infty} q^{k^2-2k}\left(q^{\lambda+\chi(m)} u\right)^{k} \nonumber \\
&&\quad+
\left( q^{n} + q^{2(n-\lfloor m/2\rfloor)} + q^{2n-\lfloor m/2\rfloor}
+ q^{2\lfloor m/2\rfloor} + q^{n+\lfloor m/2\rfloor} + q^{2n}\right)
\nonumber \\
&&\quad\times
\frac{(-q;q)_{\infty}^2}{(1-q)^2 (1-q^2)^2}
\sum_{k=0}^{\infty} q^{k^2-2k} \left(q^{\lambda+\chi(m)} u\right)^{k} \nonumber \\
&&\hspace{-0.4cm}+
\left(1+q+q^{1+n}\right)
\frac{q^{(n-\lfloor m/2\rfloor)^2}}{1-q}\left(q^{\lambda+\chi(m)} u
\right)^{-(n-\lfloor m/2\rfloor)}
\sum_{k=1}^{\infty} q^{k^2-2k} \left(q^{\lambda+\chi(m)} u\right)^{-k} \nonumber \\
&\quad&\quad+
\left( q^{n} + q^{2(n-\lfloor m/2\rfloor)} + q^{2n-\lfloor m/2\rfloor}
+ q^{2\lfloor m/2\rfloor} + q^{n+\lfloor m/2\rfloor} + q^{2n}\right)
\nonumber \\
&&\quad\times
\frac{(-q;q)_{\infty}^2}{(1-q)^2 (1-q^2)^2}
\sum_{k=1}^{\infty} q^{k^2-2k} \left(q^{\lambda+\chi(m)} u\right)^{-k}.
\label{sumup}
\end{eqnarray}
By (\ref{b4}), we obtain
\begin{eqnarray}
&&\quad  q^{n} + q^{2(n-\lfloor m/2\rfloor)} + q^{2n-\lfloor m/2\rfloor}
+ q^{2\lfloor m/2\rfloor} + q^{n+\lfloor m/2\rfloor} + q^{2n}
\nonumber \\
&&< q^{n} +q^{\tau n} + q^{n+\tau n/2}
+ q^{(2-\tau)n-2} + q^{n+ (2-\tau)n/2 -1} + q^{2n}
\nonumber \\
&&< q^{-2}\left(q^{n} +q^{\tau n} + q^{n+\tau n/2}
+ q^{(2-\tau)n} + q^{n+ (2-\tau)n/2} + q^{2n}\right)
\nonumber \\
&&< q^{-2}\left(q^{\tau n} +4 q^{n}+ q^{(2-\tau)n}\right)
\nonumber \\
&&\leq 6 q^{-2} q^{\tau n +2(1-\tau)n \I_{[1,2)}(\tau)}.
\label{b5}
\end{eqnarray}
Hence, by $1+q+q^{1+n}<3$
and $\lambda+\chi(m)\in[0,2)$, we have
\begin{eqnarray}
|r_1(n) + r_2(n)|
&<&\frac{3 q^{\lfloor m/2\rfloor^2 +2\lfloor m/2\rfloor}}
{1-q}\left(q^{\lambda+\chi(m)} u\right)^{\lfloor m/2\rfloor+1}
\sum_{k=0}^{\infty} q^{k^2-2k} \left(q^{\lambda+\chi(m)} u\right)^{k}
\nonumber \\
&\quad&\quad+
6 q^{-2} q^{\tau n +2(1-\tau)n \I_{[1,2)}(\tau)}
\frac{(-q;q)_{\infty}^2}{(1-q)^2 (1-q^2)^2}
\sum_{k=0}^{\infty} q^{k^2-2k} \left(q^{\lambda+\chi(m)} u\right)^{k}
\nonumber \\
&\quad\quad&
+\frac{3 q^{(n-\lfloor m/2\rfloor)^2}}
{1-q}\left(q^{\lambda+\chi(m)} u\right)^{-(n-\lfloor m/2\rfloor)}
\sum_{k=1}^{\infty} q^{k^2-2k} \left(q^{\lambda+\chi(m)} u\right)^{-k}
\nonumber \\
&\quad&\quad
+6 q^{-2} q^{\tau n +2(1-\tau)n \I_{[1,2)}(\tau)}
\frac{(-q;q)_{\infty}^2}{(1-q)^2 (1-q^2)^2}
\sum_{k=1}^{\infty} q^{k^2-2k}\left(q^{\lambda+\chi(m)} u\right)^{-k}
\nonumber \\
&<&\frac{3 q^{\lfloor m/2\rfloor^2 +2\lfloor m/2\rfloor}}
{1-q}u^{\lfloor m/2\rfloor+1}
\sum_{k=0}^{\infty} q^{k^2-2k} u^{k}
\nonumber \\
&\quad&\quad
+6 q^{-2} q^{\tau n +2(1-\tau)n \I_{[1,2)}(\tau)}
\frac{(-q;q)_{\infty}^2}{(1-q)^2 (1-q^2)^2}
\sum_{k=0}^{\infty} q^{k^2-2k} u^{k}
\nonumber \\
&\quad\quad&
+\frac{3 q^{(n-\lfloor m/2\rfloor)^2}}
{1-q}\left(q^{2} u\right)^{-(n-\lfloor m/2\rfloor)}
\sum_{k=1}^{\infty} q^{k^2-2k} \left(q^{2} u\right)^{-k}
\nonumber \\
&\quad&\quad
+6 q^{-2} q^{\tau n +2(1-\tau)n \I_{[1,2)}(\tau)}
\frac{(-q;q)_{\infty}^2}{(1-q)^2 (1-q^2)^2}
\sum_{k=1}^{\infty} q^{k^2-2k} \left(q^{2} u\right)^{-k}.
\nonumber \\
\end{eqnarray}
Finally, by (\ref{b4}), we have
\begin{eqnarray}
|r_1(n) + r_2(n)|
&<&\frac{3 q^{(2-\tau)^2 n^2/4 -1}}
{1-q}u^{\lfloor m/2\rfloor+1}
\sum_{k=0}^{\infty} q^{k^2-2k} u^{k}
\nonumber \\
&\quad&\quad+
6 q^{-2} q^{\tau n +2(1-\tau)n \I_{[1,2)}(\tau)}
\frac{(-q;q)_{\infty}^2}{(1-q)^2 (1-q^2)^2}
\sum_{k=0}^{\infty} q^{k^2-2k} u^{k}
\nonumber \\
&\quad\quad&
+\frac{3 q^{\tau^2 n^2 /4 -\tau n -2}}
{1-q} u^{-(n-\lfloor m/2\rfloor)}
\sum_{k=1}^{\infty} q^{k^2-4k} u^{-k}
\nonumber \\
&\quad&\quad+
6 q^{-2} q^{\tau n +2(1-\tau)n \I_{[1,2)}(\tau)}
\frac{(-q;q)_{\infty}^2}{(1-q)^2 (1-q^2)^2}
\sum_{k=1}^{\infty} q^{k^2-4k} u^{-k}
\nonumber \\
&\equiv& M(n).
\end{eqnarray}
Since, for any $\alpha >0$, $\beta >0$, and $q\in(0,1)$,
$q^{\alpha n^2} \beta^n \to 0$, as $n \to \infty$,
we have
\begin{eqnarray}
\frac{M(n)}{q^{\tau n +2(1-\tau)n \I_{[1,2)}(\tau)}}
&\longrightarrow&
6q^{-2}\frac{(-q;q)_{\infty}^2}{(1-q)^2 (1-q^2)^2}
\left[
\sum_{k=0}^{\infty} q^{k^2-2k} u^{k}
+\sum_{k=1}^{\infty} q^{k^2-4k} u^{-k}
\right],
\nonumber \\
&&\quad\quad\quad\quad\quad\quad\quad\quad\quad\quad
\quad\quad\quad\quad\quad\quad\quad\quad\quad
\quad \mbox{as $n \to \infty$}.
\label{ev}
\end{eqnarray}
Thus,
$|r_1(n) + r_2(n)|/q^{\tau n +2(1-\tau)n \I_{[1,2)}(\tau)}$
is bounded in $n \to \infty$.
Hence the proof of (\ref{order}) is completed. \qed

\end{document}